\documentclass[english]{article}
\usepackage{mathptmx}
\usepackage{helvet}

\usepackage[T1]{fontenc}
\usepackage[latin9]{inputenc}
\usepackage{amsbsy}
\usepackage{amssymb}

\makeatletter

\usepackage{geometry}

\makeatother

\usepackage{babel}

\begin{document}

\title{The Existence of Time}

\author{Joseph A. Spencer$^{\dagger}$ and James T. Wheeler$^{\ddagger}$}
\maketitle
\begin{abstract}
Of those gauge theories of gravity known to be equivalent to general
relativity, only the biconformal gauging introduces new structures
\textendash{} the quotient of the conformal group of any pseudo-Euclidean
space by its Weyl subgroup always has natural symplectic and metric
structures. Using this metric and symplectic form, we show that there
exist canonically conjugate, orthogonal, metric submanifolds if and
only if the original gauged space is Euclidean or signature 0. In
the Euclidean cases, the resultant configuration space must be Lorentzian.
Therefore, in this context, time may be viewed as a derived property
of general relativity.
\end{abstract}

\author{%
\thanks{
\author{$^{\dagger}$andy.spencer@aggiemail.usu.edu, Department of Physics,
Utah State University, Logan, Utah 84322}\maketitle
}}

\author{%
\thanks{
\author{$^{\ddagger}$jim.wheeler@usu.edu, Department of Physics, Utah State
University, Logan, Utah 84322}\maketitle
}}

\section{Introduction}

\subsection{General relativity as a biconformal gauge theory}

Since Utiyama \cite{Utiyama} wrote the first gauge theory of general
relativity, numerous authors have streamlined the procedure. Notably,
Kibble \cite{Kibble1} extended the gauge group to Poincaré, while
Ne'eman and Regge \cite{Neeman,ReggeN} (in the context of supergravity)
adapted Cartan's group quotient methods \cite{Kobayashi}. These methods
were applied extensively by Ivanov and Niederle \cite{IvanovI,Ivanov}
in the early 1980s to Poincaré, de Sitter, conformal and biconformal
gaugings. The result is that there are several ways to formulate general
relativity as a gauge theory. Presumably, one of these gaugings is
relevant to unification, if only as a low energy limit of string or
some alternative TOE.

In this paper, we develop some properties of one of these gaugings.
To begin, we briefly recall the group quotient method and its applications.
We then focus on the biconformal gauge theory, which combines the
advantages of maximum symmetry with the presence of structures not
present in other gauge theories of general relativity. We examine
all dimensions and all signatures.

A gravitational gauge theory based on a homogeneous space, $\mathcal{S}$,
of dimension $n$ may be accomplished by taking the quotient of one
of its symmetry groups \textendash{} Poincaré, inhomogeneous Weyl,
or conformal \textendash{} by a subgroup. The Maurer-Cartan connection
on the resulting fiber bundle is then generalized so that the subgroup
becomes the local symmetry group over the quotient manifold. In order
for this quotient to lead to general relativity, the appropriate subgroup
must contain the Lorentz group, and because of the balance of units
in all physical equations should probably also include dilatations.
In models where the Weyl vector is absent or pure gauge, the dilatational
symmetry may be broken by fiat. This is generally accomplished by
a global choice of the measure of time%
\footnote{S. Lloyd tells the amusing anecdote \cite{Folger}, {}``I recently
went to the National Institute of Standards and Technology in Boulder.
I said something like, \textquoteleft{}Your clocks measure time very
accurately.\textquoteright{} They told me, \textquoteleft{}Our clocks
do not measure time.\textquoteright{} I thought, Wow, that\textquoteright{}s
very humble of these guys. But they said, \textquoteleft{}No, time
is defined to be what our clocks measure.\textquoteright{}'' Indeed,
the standard second is \emph{defined} as the duration of 9,192,631,770
oscillations of the radiation from the transition between the two
hyperfine levels of the ground state of the cesium 133 atom.%
}. By allowing local dilatational symmetry it becomes possible to consider
quotients of the Weyl or conformal groups, rather than just the Poincaré/Lorentz
quotient first studied by Kibble. With a suitable choice of action,
the quotient of the inhomogeneous Weyl group by the homogeneous Weyl
group gives satisfactory formulation of general relativity. Similarly,
we can maximally extend to the conformal group. However, as noted
by Ivanov and Niederle, the quotient of the conformal group by the
inhomogeneous Weyl group \textendash{} Weyl conformal gravity \textendash{}
is a distinct theory from general relativity with a questionable low-energy
limit \cite{Flanagan} and no new structures to recommend it \cite{IvanovI,Ivanov,JWAux,JW}.
Where general relativity does emerge, along with new structures, is
from the quotient of the conformal group by the \emph{homogeneous}
Weyl group. The resulting spaces, originally called special conformal
geometries but now called biconformal spaces, are symplectic manifolds
with non-degenerate Killing metric. They admit an action linear in
the curvatures in any dimension \cite{WW}, and torsion-free solutions
reproduce general relativity on the co-tangent bundle.

It is suggestive that the natural structures \textendash{} metric
and symplectic \textendash{} are just those required by quantum theory
\cite{Berezin,Woodhouse,Curtwright,Klauder}.

Biconformal spaces are closely related to twistor space. If we introduce
a complex structure on biconformal space, we recognize it as twistor
space modified by its own additional metric and symplectic structures.

It is clear that twistor space, $\mathbb{C}^{4}$, has both Euclidean
and Lorentzian submanifolds. However, the symplectic and metric structures
of biconformal space tighten the connection between these two, giving
a nearly unique correspondence between certain Euclidean and Lorentzian
submanifolds. This unique relationship is the subject of the current
work.

\subsection{General relativistic gauge theory and the existence of time}

From the point of view of modern physics, the existence of time is
signaled by the Lorentzian symmetry of spacetime, and in particular
the signature, $s=3-1=2,$ (or, in $n$ dimensions, $s=n-2$) of the
spacetime metric. This Lorentzian metric with its invariant light
cones gives us a universal notion of causality, past, present and
future. It also gives us a pseudo-orthogonal symmetry group, $SO\left(3,1\right),$
instead of a neatly Pythagorean $SO\left(n\right).$

We present a scenario by which Lorentzian signature arises from \emph{within}
a conformal gauge theory of general relativity.

Biconformal gauging of the conformal group gives rise to both a natural
metric and a natural symplectic structure. We gauge an arbitrary,
$p+q=n$ dimensional pseudo-Euclidean space $S_{p,q}$ of signature
$s_{0}=p-q,$ to produce the corresponding biconformal geometry. Our
central theorem states that orthogonal, canonically conjugate, metric
submanifolds of biconformal space exist if and only if the original
$n$-dim space is Euclidean $\left(s_{0}=\pm\, n\right),$ or of vanishing
signature $\left(s_{0}=0\right)$. The signature of the induced metric
on the configuration space is $n-2$ (Lorentzian) or $2,$ respectively.
When $n=4$ or $n=2m+1,$ the configuration space is always a Lorentzian
spacetime.

Our presentation proceeds as follows. Section 2 begins with a description
of the zero-curvature case of biconformal gauge theory. This is sufficient
for a general proof. We then introduce the conformal Killing metric
and its restriction to biconformal space, describe the natural symplectic
form, and make the definitions required to state our central theorem.
The Section concludes with our statement of the central (Signature)
theorem. In Section 3, we find the general form for an orthogonal,
canonically conjugate basis and in Section 4 we impose the condition
of conformal flatness of the momentum submanifold. Then, in order
to establish the \emph{uniqueness} the conclusions, we must completely
solve the biconformal structure equations. This somewhat lengthy calculation
is described in Section 5. The final conclusions of the Signature
Theorem rest on an analysis, in Section 6, of the consistency of the
signature of the induced metric on both the configuration and momentum
submanifolds. We conclude, in Section 7, with a brief summary of our
results.

\section{The Signature Theorem}

In this Section, we provide basic definitions required for the statement
of our central theorem, followed by a statement of the theorem and
a brief outline of the proof. The remaining Sections provide details
of the proof.

\subsection{The construction of biconformal space}

Let $\mathcal{S}_{p,q}^{n}$ be a pseudo-Euclidean space of signature
$s_{0}=p-q$ and dimension $n=p+q.$ Let the $SO\left(p,q\right)$-invariant
metric on this space be\begin{equation}
\eta_{ab}=diag\left(1,\ldots,1,-1,\ldots,-1\right)\label{Defining metric}\end{equation}
with inverse $\eta^{ab}.$ Compactify the space by appending a point
at infinity for every null vector from the origin. For Euclidean space,
$p=\pm\, n,$ this will be a single point inverse to the origin; for
Minkowski space, $p=n-1,$ a single light cone is required. With the
exception of the Euclidean cases, the null subspace is of dimension
$n-1$.

We may now define the conformal group of $\mathcal{S}_{p,q}$ as the
set of transformations which preserve the metric, or equivalently
the infinitesimal line element, $ds^{2}=\eta_{ab}dx^{a}dx^{b}$, up
to an overall factor. The conformal transformations include $SO\left(p,q\right)$
transformations, translations, dilatations, and special conformal
transformations; together these make up $SO\left(p+1,q+1\right).$

The Lie algebra $so\left(p+1,q+1\right)$ may be written in terms
of basis $1$-forms as the Maurer-Cartan structure equations,\begin{equation}
\mathbf{d\omega}^{\Delta}=-\frac{1}{2}c_{\Sigma\Lambda}^{\quad\;\Delta}\mathbf{\omega}^{\Sigma}\mathbf{\omega}^{\Lambda}\label{Maurer-Cartan}\end{equation}
where $c_{\Sigma\Lambda}^{\quad\;\Delta}$ are the structure constants
of the conformal group and $\mathbf{\omega}^{\Delta}\in\left\{ \mathbf{\omega}_{b}^{a}=-\eta^{ac}\eta_{bd}\mathbf{\omega}_{c}^{d},\mathbf{\omega}^{a},\mathbf{\omega}_{b},\mathbf{\omega}\right\} $,
where capital Greek indices run over all $\frac{\left(n+1\right)\left(n+2\right)}{2}$
dimensions of $SO\left(p+1,q+1\right),$ and lowercase Latin indices
take values of $1$ to $n$. These forms are dual to the generators
of $SO\left(p+1,q+1\right):$ $\mathbf{\omega}_{b}^{a}$ are dual
to the generators of $SO\left(p,q\right),$ $\omega^{a}$ are dual
to the generators of translations, $\omega_{b}$ are dual to the generators
of special conformal transformations, and $\omega$ is dual to the
generator of dilatations. Expanding eq.(\ref{Maurer-Cartan}),\begin{eqnarray}
\mathbf{d\omega}_{b}^{a} & = & \mathbf{\omega}_{b}^{c}\mathbf{\omega}_{c}^{a}+2\Delta_{db}^{ac}\mathbf{\omega}_{c}\mathbf{\omega}^{d}\label{Lorentz}\\
\mathbf{d\omega}^{a} & = & \mathbf{\omega}^{c}\mathbf{\omega}_{c}^{a}+\mathbf{\omega\omega}^{a}\label{Solder form}\\
\mathbf{d\omega}_{a} & = & \mathbf{\omega}_{a}^{c}\mathbf{\omega}_{c}+\mathbf{\omega}_{a}\mathbf{\omega}\label{Co-solder}\\
\mathbf{d\omega} & = & \mathbf{\omega}^{a}\mathbf{\omega}_{a}\label{Dilatation}\end{eqnarray}
where we define $\Delta_{db}^{ac}\equiv\frac{1}{2}\left(\delta_{d}^{a}\delta_{b}^{c}-\eta_{bd}\eta^{ac}\right).$
These are the Maurer-Cartan equations for the conformal group with
respect to $\eta_{ab}$. They are completely equivalent to the Lie
algebra commutation relations, with the Jacobi identity following
from the integrability condition $\mathbf{d}^{2}=0$.
\begin{description}
\item [{Definition}] A flat biconformal space is the quotient of the conformal
group by its Weyl subgroup (i.e., $SO\left(p,q\right)$ and dilatations).
It is a principal fiber bundle with Weyl local symmetry group and
$2n$-dim base manifold \cite{Ivanov,JW,WW}.
\end{description}
The properties we study below, including metric, symplectic structure,
and submanifolds, all generalize directly to curved biconformal spaces
\cite{WW}. Our results therefore apply immediately to this more general
class of spaces by continuity; in particular it has been shown that
generic, torsion-free biconformal spaces still retain the symplectic
form. Since biconformal spaces have local dilatational symmetry, definitions
and theorems that would refer to a manifold being flat (i.e., vanishing
Riemann tensor) are generalized to mean that the Weyl curvature tensor
vanishes. Then there exists a conformal gauge in which the space is
flat.

Flat biconformal spaces are described by eqs.(\ref{Lorentz}-\ref{Dilatation}),
with the connection forms $\mathbf{\omega}_{b}^{a},\mathbf{\omega}$
taken to be horizontal, i.e., expandable in terms of the basis forms
$\mathbf{\omega}^{a}$ and $\mathbf{\omega}_{a}.$ Solutions to these
equations are given elsewhere \cite{IvanovI,JW,WW}; however we present
a new solution satisfying certain conditions described in the next
Section. The generality of this new solution is important in establishing
the uniqueness property of our central result.

Now we turn to a discussion of the Killing metric and symplectic structure
of the conformal group, and a precise statement of our central theorem.

\subsection{Natural metric and symplectic structure}

The Killing metric of the conformal group is built from the structure
constants $c_{\Sigma\Lambda}^{\quad\quad\Delta}$ as the symmetric
form\[
K_{\Sigma\Lambda}=\frac{1}{2n}c_{\Delta\Sigma}^{\quad\quad\Theta}c_{\Theta\Lambda}^{\quad\quad\Delta}=\left(\begin{array}{cccc}
\frac{1}{2}\Delta_{db}^{ac} & 0 & 0 & 0\\
0 & 0 & \delta_{b}^{a} & 0\\
0 & \delta_{a}^{b} & 0 & 0\\
0 & 0 & 0 & 1\end{array}\right)\]
where $\frac{1}{2}\Delta_{db}^{ac}$ corresponds to $SO\left(p,q\right)$
transformations, the middle double block corresponds to translations
and special conformal transformations, and the final $1$ in the lower
right to dilatations. Notice that, of the gauge theories discussed
in the introduction, the biconformal case is the only one where the
restriction of $K_{\Sigma\Lambda}$ to the base manifold is non-degenerate
\textendash{} when $K_{\Sigma\Lambda}$ above is restricted to the
biconformal manifold it becomes $\left(\begin{array}{cc}
0 & \delta_{b}^{a}\\
\delta_{a}^{b} & 0\end{array}\right)$. We therefore have a natural metric on both the full group manifold
and on the $2n$-dim biconformal space.

In addition to the metric, even curved biconformal spaces generically
have a natural symplectic structure. Since $\left(\mathbf{\omega}^{a},\mathbf{\omega}_{a}\right)$
together span the base manifold, the dilatational structure equation\[
\mathbf{d\omega}=\mathbf{\omega}^{a}\mathbf{\omega}_{a}\]
is necessarily a closed, non-degenerate two form.

We now turn to the question of when biconformal spaces coincide with
our usual notions of a relativistic phase space. We begin with the
definition \cite{AbrahamAndMarsden},
\begin{description}
\item [{Definition}] (Abraham, Marsden) A \textbf{phase space} is a symplectic
manifold which is the cotangent bundle of a Riemannian or pseudo-Riemannian
manifold.
\end{description}
It has been shown that all torsion-free biconformal spaces solving
the curvature-linear field equations are cotangent bundles \cite{WW}.
The central issue relating biconformal and phase spaces is the metric.
The most interesting feature of our proof hinges on the \textit{signatures}
of the induced metrics. This is why we only need to consider flat
biconformal space.

There is an important difference between the metric of a phase space
and the metric of a biconformal space. The use of a Riemannian manifold
in the Abraham-Marsden definition implies the presence of a metric
on the configuration manifold. This induces an inner product on the
cotangent spaces, so the metric on the momentum submanifolds is inverse
(since $p_{a}$ is covariant) to the flat form of the configuration
metric. There is \textit{not}, in general, a single metric on the
entire phase space.

By contrast, there \textit{is} a metric on the entirety of a biconformal
space. The Killing metric of the conformal group, restricted to the
biconformal submanifold, is nondegenerate. This means that if the
biconformal space is identified with a phase space, the configuration
and momentum metrics are \textit{a priori} independent. Nonetheless,
the proof below shows that in the majority of cases the induced metric
on the configuration space is uniquely Lorentzian, and the metric
on the momentum spaces is the \textit{negative} of the inverse to
the corresponding flat Lorentz metric. The existence of time may be
attributed to the necessity for the Lorentz signature of the configuration
submanifold. Our concluding remarks focus on this point. It has been
argued \cite{YQMisC,QM} that the imaginary unit in Dirac's quantization
rule, replacing Poisson brackets by $-\frac{i}{\hbar}$ times the
commutator, may be attributed to the relative minus sign between the
configuration and momentum space metrics.

In order to precisely relate the biconformal metric to the configuration
space metric, we make the following definition:
\begin{description}
\item [{Definition}] A \textbf{metric phase space} is a phase space with
metric, having a basis $\left(\mathbf{\chi}^{a},\mathbf{\eta}_{b}\right)$
such that the following conditions hold:\end{description}
\begin{enumerate}
\item $\mathbf{\chi}^{a}$ and $\mathbf{\eta}_{b}$ are canonically conjugate.
\item $\mathbf{\chi}^{a}$ and $\mathbf{\eta}_{b}$ are orthogonal with
respect to the Killing metric,\begin{equation}
\left\langle \mathbf{\chi}^{a},\mathbf{\eta}_{b}\right\rangle =0\label{Orthogonality}\end{equation}
while the induced configuration space metric, \begin{equation}
h^{ab}\equiv\left\langle \mathbf{\chi}^{a},\mathbf{\chi}^{b}\right\rangle \label{Nondegenerate metric}\end{equation}
is non-degenerate. It follows that the momentum space metric is also
non-degenerate. Orthogonality is required so that the configuration
metric is well-defined.
\item $\mathbf{\chi}^{a}$ and $\mathbf{\eta}_{b}$ are separately involute.
Thus, the conditions $\mathbf{\chi}^{a}=0$ and $\mathbf{\eta}_{a}=0$
each provide a projection to an $n$-dim metric submanifold. The first
is called momentum space; the second is called configuration space. 
\end{enumerate}
Since a metric phase space is a phase space, it is an even dimensional
manifold, $\mathcal{M}$, with symplectic structure. Since it must
be a cotangent bundle, the momentum submanifold must be flat. We immediately
have the following lemma.
\begin{description}
\item [{Lemma$\:$1}] \textit{There is a conformal gauge in which flat
biconformal space is a metric phase space if there exists a basis,}
$\left(\mathbf{\chi}^{a},\mathbf{\eta}_{b}\right)$\textit{, such
that the following conditions hold:}\end{description}
\begin{enumerate}
\item In terms of $\left(\mathbf{\chi}^{a},\mathbf{\eta}_{b}\right),$ eq.(\ref{Dilatation})
takes the form $\mathbf{d\omega}=\mathbf{\chi}^{a}\mathbf{\eta}_{a}.$
\item $\mathbf{\chi}^{a}$ and $\mathbf{\eta}_{b}$ are orthogonal with
respect to the Killing metric of the conformal group, i.e., $\left\langle \mathbf{\chi}^{a},\mathbf{\eta}_{b}\right\rangle _{K}=0$.
\item The structure equations for $\mathbf{\chi}^{a}$ and $\mathbf{\eta}_{b}$
are separately involute.
\item The Weyl curvature of the momentum submanifold vanishes. \end{enumerate}
\begin{description}
\item [{Proof}] Since $\mathbf{d\omega}$ is the symplectic form, condition
1 holds if and only if $\mathbf{\chi}^{a}$ and $\mathbf{\eta}_{b}$
are canonically conjugate. Since the Killing metric is nondegenerate,
and $\left\langle \mathbf{\chi}^{a},\mathbf{\eta}_{b}\right\rangle =0,$
the inner products $g^{ab}\equiv\left\langle \mathbf{\chi}^{a},\mathbf{\chi}^{b}\right\rangle $
and $g_{ab}^{\prime}\equiv\left\langle \mathbf{\eta}_{a},\mathbf{\eta}_{b}\right\rangle $
are necessarily non-degenerate. Condition 3 is unchanged from the
definition of a metric phase space. Finally, condition 4 guarantees
the existence of a gauge in which the momentum submanifold is flat.
\end{description}
These conditions on biconformal space produce submanifolds which can
be identified with the configuration manifold and cotangent spaces
of phase space. We include the possibility of arbitrary signature
for the metric submanifolds, though we will have considerably more
to say about this below.

\subsection{The Signature Theorem}

Now we come to the main theorem of this paper, and two corollaries.
The full theorem applies to any initial dimension and signature:
\begin{description}
\item [{Theorem}] \textit{\emph{(Signature Theore}}\textit{m}\textit{\emph{)}}\textit{
Flat} $2n$\textit{-dim (}$n>2$\textit{) biconformal space is a metric
phase space if and only if the signature,} $s,$\textit{ of} $\eta_{ab}$\textit{
is} $\pm n$\textit{ or} $0.$\textit{ These three possibilities lead
to the following signatures for the configuration submanifold:}\[
\begin{array}{cc}
s & s_{configuration~space}\\
n & n-2\ \left(Lorentz\right)\\
-n & n-2\ \left(Lorentz\right)\\
0\  & -2\end{array}\]
\textit{The signature of the momentum submanifold is always the negative
of the signature of the configuration submanifold. Since we can have}
$s=0$\textit{ only if} $n$\textit{ is even, configuration space
is uniquely Lorentz if} $n$\textit{ is odd.}
\end{description}
Since, when $n=4,$ the signature $-2$ is also Lorentzian, we have
the immediate corollary,
\begin{description}
\item [{Corollary$\:$1}] \textit{Flat} $8$\textit{-dim biconformal space
reduces to a metric phase space if and only if the initial} $4$\textit{-dim
space we gauge, }$\mathcal{S}_{p,q}^{n=4}$,\textit{ is Euclidean
or signature zero, and the resulting configuration space is necessarily
Lorentzian.}
\end{description}
Since odd-dimensional spaces cannot have zero signature, we also have,
\begin{description}
\item [{Corollary$\:$2}] \textit{A biconformal space built by gauging
any odd-dimensional space, }$\mathcal{S}_{p,q}^{n=2m+1}$,\textit{
reduces to a metric phase space if and only if the initial space is
Euclidean, $s_{0}=\pm n$. The resulting configuration space is necessarily
Lorentzian.}
\end{description}
\bigskip{}

The details of the proof lead to the following additional conclusions:
\begin{enumerate}
\item There exists a basis $\left(\mathbf{\chi}^{a},\mathbf{\eta}_{b}\right)$
such that the connection takes the form\begin{eqnarray*}
\mathbf{\omega}_{b}^{a} & = & 2\Delta_{db}^{ac}\left(y_{c}-d_{c}\right)\mathbf{d}v^{d}+\frac{2}{y^{2}}\Delta_{db}^{ac}\eta^{de}y_{e}\mathbf{\eta}_{c}\\
\mathbf{\chi}^{a} & = & \mathbf{d}v^{a}\\
\mathbf{\eta}_{a} & = & \mathbf{d}y_{a}+\left(y_{b}d_{a}+y_{a}d_{b}-\eta_{ab}\left(\eta^{ef}y_{e}d_{f}\right)\right)\mathbf{d}v^{b}\\
\mathbf{\omega} & = & -y_{a}\mathbf{d}v^{a}-\frac{1}{y^{2}}\eta^{ab}y_{a}\mathbf{d}y_{b}\end{eqnarray*}
where $\eta_{ab}$ is given by eq.(\ref{Defining metric}), and where
either $d_{a}=\frac{\eta_{ab}v^{b}}{v^{2}}$ or $d_{a}=\frac{c_{a}}{a_{0}+c_{a}v^{a}}$
where $a_{0}$ is a constant and $c_{a}$ is a constant null vector
in the original metric, $\eta^{ab}c_{a}c_{b}=0$. The coordinates
$v^{a}$ and $y_{a}$ are defined in the proof.
\item In the $\left(\mathbf{\chi}^{a},\mathbf{\eta}_{a}\right)$ basis,
the Killing metric is given by\begin{eqnarray*}
\left\langle \mathbf{\chi}^{a},\mathbf{\chi}^{b}\right\rangle  & = & -h^{ab}\\
\left\langle \mathbf{\chi}^{a},\mathbf{\eta}_{a}\right\rangle  & = & 0\\
\left\langle \mathbf{\eta}_{a},\mathbf{\eta}_{b}\right\rangle  & = & h_{ab}\end{eqnarray*}
where $h_{ab}=2y_{a}y_{b}-y^{2}\eta_{ab}$.\end{enumerate}
\begin{description}
\item [{Proof\,(outline)}] The proof of the Signature Theorem, comprising
the bulk of the remaining Sections, is accomplished by imposing the
conditions of the Lemma $1$ on a general flat biconformal space.
Here we outline the sequence of demonstrations:\end{description}
\begin{enumerate}
\item (Section 3) Using the natural symplectic form and the Killing metric,
impose conditions 1 and 2 of the Lemma on a general linear transformation
between bases $\left(\mathbf{\chi}^{a},\mathbf{\eta}_{b}\right)$
and $\left(\mathbf{\omega}^{a},\mathbf{\omega}_{b}\right)$. Up to
transformations within each subspace, this gives an expansion of the
basis $\left(\mathbf{\chi}^{a},\mathbf{\eta}_{b}\right)$ in terms
of the original basis $\left(\mathbf{\omega}^{a},\mathbf{\omega}_{b}\right)$
and one additional matrix, $h^{ab}$. Recasting the Maurer-Cartan
structure equations in the new basis $\left(\mathbf{\chi}^{a},\mathbf{\eta}_{b}\right)$,
we note the form of the involution conditions required by condition
3 of Lemma 1.
\item (Section 4) Prove that conformal flatness of the momentum subspace
(Lemma 1, condition 3) implies\[
h^{ab}=\frac{\left(n-2\right)h}{u^{2}}\left(-2u^{a}u^{b}+u^{2}\eta^{ab}\right)\]
for some vector field $u^{a}$.
\item (Section 5) To arrive at a final form for $h^{ab}$ and prove that
it gives a solution for the biconformal space, we completely solve
the structure equations, eqs.(\ref{Lorentz}-\ref{Dilatation}). First,
setting $\mathbf{\chi}^{a}=0$, we solve for the restriction of the
connection to the momentum submanifold. Making suitable gauge choices,
we find a distinct form for the inverse metric $h^{ab}$,\[
h^{ab}=e^{-2\sigma}\left(-2\sigma^{,ab}+2\sigma^{,a}\sigma^{,b}-\sigma^{,c}\sigma^{,d}\eta_{cd}\eta^{ab}\right)\]
depending on a single function $\sigma$ and its $y_{a}$-derivatives.
Throughout, we denote derivatives with respect to $v^{a}$ and $y_{a}$
by $\frac{\partial f}{\partial v^{a}}=f_{,a}$ and $\frac{\partial f}{\partial y_{a}}=f^{,a}$,
respectively. We then relate the two forms of $h^{ab}$ found in Sections
2 and 3. This shows that $u^{a}$ must be a gradient with respect
to $y_{a},$ $u^{a}=\frac{\partial u}{\partial y_{a}},$ for some
function $u,$ where $u$ and $\sigma$ must satisfy the coupled differential
equations:\begin{eqnarray*}
\sigma^{,ab} & = & \sigma^{,b}\sigma^{,a}+\beta u^{a}u^{b}-\frac{1}{2}\left(\sigma^{2}+\beta u^{2}\right)\eta^{ab}\\
u^{,ab} & = & u^{,a}\sigma^{,b}+u^{b}\sigma^{,a}-\left(u_{c}\sigma^{,c}\right)\eta^{ab}\end{eqnarray*}
After solving these equations to complete the solution on the momentum
submanifold, we extend the result back to the full biconformal space
and solve the full structure equations. The results prove the claims
following the Corollary, providing the form of the connection and
the unique form of the induced configuration space metric.
\item (Section 6) We study the signature of the Killing metric on both the
configuration and momentum submanifolds. Imposing consistency of the
signature across the cotangent bundle proves the last statement in
the theorem. 
\end{enumerate}
We end with a brief discussion of the physical meaning of the results.

\section{Orthogonal symplectic bases}

We begin by expressing the biconformal structure equations in an orthogonal
symplectic basis. A general basis is given by\begin{eqnarray*}
\mathbf{\chi}^{a} & = & A_{\quad b}^{a}\mathbf{\omega}^{b}+B^{ab}\mathbf{\omega}_{b}\\
\mathbf{\eta}_{a} & = & C_{a}^{\quad b}\mathbf{\omega}_{b}+D_{ab}\mathbf{\omega}^{b}\end{eqnarray*}
We demand two conditions. First, the inner products must be non-degenerate
and orthogonal, according to eqs. (\ref{Nondegenerate metric}) and
(\ref{Orthogonality}). Once we guarantee orthogonality, the nondegeneracy
of the Killing metric insures that $g^{ab}$ and $\left\langle \mathbf{\eta}_{a},\mathbf{\eta}_{b}\right\rangle =g_{ab}^{\prime}$
are non-degenerate. Second, we require the basis to be canonical,
$\mathbf{d\omega}=\mathbf{\omega}^{a}\mathbf{\omega}_{a}=\mathbf{\chi}^{a}\mathbf{\eta}_{a}$.
Substituting, and recasting all three orthogonality and three conjugacy
conditions in matrix notation, we find the conditions,\begin{eqnarray}
g & = & BA^{t}+AB^{t}\label{OrthoNor1}\\
0 & = & AC^{t}+BD^{t}\label{OrthoNor2}\\
g^{\prime} & = & CD^{t}+DC^{t}\label{OrthoNor3}\\
0 & = & A^{t}D-D^{t}A\label{OrthoNor4}\\
\mathbf{1} & = & A^{t}C-D^{t}B\label{OrthoNor5}\\
0 & = & B^{t}C-C^{t}B\label{OrthoNor6}\end{eqnarray}
where $g$ and $g^{\prime}$ are non-degenerate and symmetric.

To solve these equations, multiply eq.(\ref{OrthoNor1}) equation
by $D^{t},$ then use eq.(\ref{OrthoNor4}) to show that $D^{t}g=D^{t}BA^{t}+A^{t}DB^{t}$.
Next use the transpose of eq.(\ref{OrthoNor2}) to replace $DB^{t},$
and finally eq.(\ref{OrthoNor5}). to arrive at $D=-g^{-1}A$. Similarly,
we find that $C=g^{-1}B$. Substituting these results shows that $g^{\prime}=-g^{-1}$
and reduces the full system to \begin{eqnarray*}
g & = & 2BA^{t}=g^{t}\\
\mathbf{1} & = & 2A^{t}g^{-1}B\end{eqnarray*}
The non-degeneracy of $g$ now implies the non-degeneracy of $A$
and $B.$ Therefore we may solve for $B$ to get $B=\frac{1}{2}g\left(A^{t}\right)^{-1}$.
The final equation is then identically satisfied.

Now define a new matrix, $h^{ab}$, by $h=A^{-1}g\left(A^{t}\right)^{-1}$,
and denote its inverse by $h_{ab}$. Then up to a change of basis
within the $\mathbf{\chi}^{a}$ and $\mathbf{\eta}_{a}$ submanifolds
(by $A_{\quad b}^{a}$ and $\left(A^{-1}\right)_{\quad a}^{b}$, respectively),
the most general orthogonal, symplectic basis is 

\begin{eqnarray*}
\mathbf{\chi}^{a} & = & \mathbf{\omega}^{b}+\frac{1}{2}h^{bc}\mathbf{\omega}_{c}\\
\mathbf{\eta}_{a} & = & \frac{1}{2}\mathbf{\omega}_{a}-h_{ab}\mathbf{\omega}^{b}\\
\mathbf{\omega}^{a} & = & \frac{1}{2}\left(\mathbf{\chi}^{a}-h^{ab}\mathbf{\eta}_{b}\right)\\
\mathbf{\omega}_{a} & = & \mathbf{\eta}_{a}+h_{ab}\mathbf{\chi}^{b}\end{eqnarray*}

In terms of these, direct substitution into the structure equations,
eqs.(\ref{Lorentz}-\ref{Dilatation}), yields\begin{eqnarray}
\mathbf{d\omega}_{b}^{a} & = & \mathbf{\omega}_{b}^{c}\mathbf{\omega}_{c}^{a}+\Delta_{db}^{ac}\left(\mathbf{\eta}_{c}+h_{cf}\mathbf{\chi}^{f}\right)\left(\mathbf{\chi}^{d}-h^{de}\mathbf{\eta}_{e}\right)\label{Chi/Eta Lorentz}\\
\mathbf{d\chi}^{a} & = & \mathbf{\chi}^{c}\mathbf{\omega}_{c}^{a}+\mathbf{\omega\chi}^{a}+\frac{1}{2}\mathbf{D}h^{ac}\left(\mathbf{\eta}_{c}+h_{cd}\mathbf{\chi}^{d}\right)\label{Chi}\\
\mathbf{d\eta}_{a} & = & \mathbf{\omega}_{a}^{b}\mathbf{\eta}_{b}-\mathbf{\omega\eta}_{a}-\frac{1}{2}\mathbf{D}h_{ab}\left(\mathbf{\chi}^{b}-h^{bc}\mathbf{\eta}_{c}\right)\label{Eta}\\
\mathbf{d\omega} & = & \mathbf{\chi}^{a}\mathbf{\eta}_{a}\label{Chi/Eta Weyl}\end{eqnarray}
where we have written the results in terms of the $SO\left(p,q\right)$-
and Weyl-covariant derivatives of $h_{ab}$ and $h^{ab}$,\begin{eqnarray*}
\mathbf{D}h^{ab} & = & \mathbf{d}h^{ab}+h^{cb}\mathbf{\omega}_{c}^{a}+h^{ac}\mathbf{\omega}_{c}^{b}-2h^{ab}\mathbf{\omega}\\
\mathbf{D}h_{ab} & = & \mathbf{d}h_{ab}-h_{cb}\mathbf{\omega}_{a}^{c}-h_{ac}\mathbf{\omega}_{b}^{c}+2h_{ab}\mathbf{\omega}\end{eqnarray*}
Eqs.(\ref{Chi/Eta Lorentz} - \ref{Chi/Eta Weyl}) describe the connection
1-forms, $\mathbf{\omega}_{b}^{a}$ and $\mathbf{\omega},$ and the
basis forms, $\left(\mathbf{\chi}^{a},\mathbf{\eta}_{a}\right)$,
of the spaces we wish to study. Our goal, over the next few Sections,
is to solve these equations subject to two further conditions required
by Lemma 1. According to condition 3, both $\mathbf{\chi}^{a}$ and
$\mathbf{\eta}_{a}$ must be in involution. Also, in order for there
to exist a conformal gauge in which the biconformal space is a co-tangent
bundle, the momentum submanifold must be conformally flat. We end
this Section with a discussion of the involution conditions, then
take up conformal flatness and related conditions in Section 4.

In order for $\mathbf{\chi}^{b}=0$ and $\mathbf{\eta}_{b}=0$ to
specify submanifolds, each of the corresponding structure equations,
(\ref{Chi}) and (\ref{Eta}) must be in involution. Therefore, we
demand\begin{eqnarray}
\left.\mathbf{D}h^{ac}\mathbf{\eta}_{c}\right|_{\mathbf{\chi}^{a}=0} & = & 0\label{Chi Involution}\\
\left.\mathbf{D}h_{ab}\mathbf{\chi}^{b}\right|_{\mathbf{\eta}_{c}=0} & = & 0\label{Eta Involution}\end{eqnarray}
These conditions will be satisfied by assuming they hold and using
the resulting two involutions to study the momentum submanifold separately.
Imposing conformal flatness places strong constraints on the corresponding
part of the connection. Then, extending back to the full biconformal
space in Section 5, we find the full connection. This procedure leads,
in Subsection 5.3.3, to final forms for the metric, connection forms,
and basis forms that solve eqs.(\ref{Chi/Eta Lorentz} - \ref{Chi/Eta Weyl}).
These final forms are easily checked to solve the involution conditions,
eqs.(\ref{Chi Involution}) and (\ref{Eta Involution}), thereby showing
that the solution provides necessary and sufficient conditions for
involution.

\section{Conformal flatness of the momentum submanifold}

As a first step in solving the structure equations, we study the solution
of the reduced set of structure equations describing the momentum
submanifold, which arises from the involution of $\mathbf{\chi}^{a}.$

To start, we assume $\mathbf{\chi}^{a}$ is in involution. By the
Frobenius theorem, there exist $n$ coordinates $v^{a}$ such that
$\mathbf{\chi}^{a}=\chi_{\beta}^{\quad a}\mathbf{d}v^{\beta}.$ Holding
$v^{a}$ constant then restricts to submanifolds described by setting
$\mathbf{\chi}^{a}=0$ in eqs.(\ref{Chi/Eta Lorentz} - \ref{Chi/Eta Weyl}).
This gives\begin{eqnarray}
\mathbf{d}\boldsymbol{\beta}_{b}^{a} & = & \boldsymbol{\beta}_{b}^{c}\boldsymbol{\beta}_{c}^{a}-\Delta_{db}^{ac}h^{de}\boldsymbol{\eta}_{c}\boldsymbol{\eta}_{e}\label{Reduced Lorentz}\\
\mathbf{d\boldsymbol{\eta}}_{a} & = & \boldsymbol{\beta}_{a}^{b}\boldsymbol{\eta}_{b}-\boldsymbol{\tau}\boldsymbol{\eta}_{a}\label{Reduced Eta}\\
\mathbf{d}\boldsymbol{\tau} & = & 0\label{Reduced Dilatation}\end{eqnarray}
where $\mathbf{\beta}_{b}^{a}\equiv\left.\mathbf{\omega}_{b}^{a}\right\vert _{\mathbf{\chi}^{a}=0}$
and $\mathbf{\tau}=\left.\mathbf{\omega}\right\vert _{\mathbf{\chi}^{a}=0}$
may be expanded in terms of $\mathbf{\eta}_{a}$ only. We also have
the metric, $h_{ab}=-\left\langle \mathbf{\eta}_{a},\mathbf{\eta}_{b}\right\rangle $
and the constraint condition,\[
0=\left.\mathbf{D}h^{ac}\right\vert _{\mathbf{\chi}^{a}=0}\mathbf{\eta}_{c}=h^{ac;b}\mathbf{\eta}_{b}\mathbf{\eta}_{c}\]
Note that \textit{on the submanifold} the constraint follows automatically
from the structure equation for $\mathbf{\eta}_{a}$, eq.(\ref{Reduced Eta}),
since the covariant constancy of the basis implies the constancy of
the associated metric, $\mathbf{D}_{\left(\mathbf{\beta}\right)}h^{ac}\left(y\right)=0$.
A similar set of equations, found by setting $\mathbf{\eta}_{a}=0,$
describes the configuration submanifold, including $\mathbf{D}_{\left(\mathbf{\alpha}\right)}h^{ac}\left(v\right)=0$.
This must not be construed to mean that the full covariant derivative
of the metric, $\mathbf{D}_{\left(\mathbf{\alpha}+\mathbf{\beta}\right)}h^{ac}\left(y,v\right)$
vanishes. It does not.

Eqs.(\ref{Reduced Lorentz} - \ref{Reduced Dilatation}) may be interpreted
as those of a Riemannian geometry if we choose a gauge where $\mathbf{\tau}=0$,
or trivial Weyl geometry if $\mathbf{\tau}=\mathbf{d}\phi$. In either
case, the solder form is $\mathbf{\eta}_{a},$ the spin connection
is $\mathbf{\beta}_{c}^{a},$ and the curvature $2$-form is\begin{equation}
\mathbf{R}_{b}^{a}=\mathbf{d\beta}_{b}^{a}-\mathbf{\beta}_{b}^{c}\mathbf{\beta}_{c}^{a}=-\Delta_{db}^{ac}h^{de}\mathbf{\eta}_{c}\mathbf{\eta}_{e}\label{Submanifold curvature}\end{equation}
We also have the inverse metric, $h_{ab}=-\left\langle \mathbf{\eta}_{a},\mathbf{\eta}_{b}\right\rangle .$

\subsection{Bianchi identities and the antisymmetry of the curvature}

For the interpretation as a Riemannian geometry to be valid, eq.(\ref{Submanifold curvature})
must satisfy the Bianchi identities of eqs.(\ref{Reduced Lorentz}
- \ref{Reduced Dilatation}). In addition, if $h^{ab}$ is to be the
metric then it must preserve the antisymmetry of $\mathbf{R}_{b}^{a},$
that is, $\mathbf{R}_{b}^{a}=-h_{bc}h^{ad}\mathbf{R}_{d}^{c}$.

We first check the Bianchi identities. These follow by taking the
exterior derivative of eqs.(\ref{Reduced Lorentz}-\ref{Reduced Dilatation})
and using the Poincarè lemma, $\mathbf{d}^{2}=0.$ This guarantees
that the equations are integrable. The conditions are satisfied identically
for the Weyl vector, $\boldsymbol{\tau}$, and the solder form, $\boldsymbol{\eta}_{a}$.
Taking the exterior derivative of eq.(\ref{Reduced Lorentz}) for
the $SO\left(p,q\right)$ spin connection and substituting for $\mathbf{d}\boldsymbol{\beta}_{a}^{b}$
in the result leads to the condition\[
0=\left(-\delta_{b}^{f}h^{ag;e}+\delta_{b}^{g}h^{af;e}-\eta^{af}\eta_{bd}h^{dg;e}+\eta^{ag}\eta_{bd}h^{df;e}\right)\mathbf{\eta}_{e}\mathbf{\eta}_{f}\mathbf{\eta}_{g}\]
where the semicolon is defined by the restricted covariant derivative
as $\left.\mathbf{D}h^{ab}\right\vert _{\mathbf{\chi}_{c}=0}\equiv h^{ab;c}\mathbf{\eta}_{c}$.
Defining the combination $v^{c}\equiv\eta_{de}h^{d\left[e;c\right]}$
and taking the trace on $fb$ shows that\begin{equation}
h^{a\left[b;c\right]}=\frac{1}{n-1}\left(\eta^{ab}v^{c}-\eta^{ac}v^{b}\right)\label{Spin Bianchi}\end{equation}
This has no further nonvanishing trace. Substituting this condition
into the original identity satisfies the full equation, so eq.(\ref{Spin Bianchi})
is the necessary and sufficient condition on $h^{ab}$ from the $SO\left(p,q\right)$
Bianchi identity. The involution condition, however, is $\left.\mathbf{D}h^{ac}\mathbf{\eta}_{c}\right|_{\mathbf{\chi}^{a}=0}=0$
so on the $\mathbf{\chi}^{a}=0$ submanifold, we have the stronger
condition $h^{ab;c}-h^{ac;b}=0$, and eq.(\ref{Spin Bianchi}) is
satisfied.

The next condition we observe that $h_{ab}$ is constrained by the
antisymmetry of the curvature. Demanding $h_{bc}h^{ad}\mathbf{R}_{d}^{c}=-\mathbf{R}_{b}^{a}$
leads to the condition,\[
h_{bf}\eta^{fc}\left(\eta_{gd}h^{ga}h^{de}\right)-h_{bf}\eta^{fe}h^{ag}\eta_{gd}h^{dc}=\eta^{ae}\eta_{bd}h^{dc}-\eta^{ac}\eta_{bd}h^{de}\]
This equation holds if and only if its trace equation,\begin{equation}
h^{ab}=\lambda\eta^{ac}\eta^{bd}h_{cd}\label{h for antisym}\end{equation}
holds, where $\lambda=\frac{\eta_{cd}h^{cd}}{h_{bc}\eta^{bc}}.$ Thus,
the inverse of the original metric, $\eta^{ab},$ must relate the
metric $h^{ab}$ to its inverse, $h_{ab}.$

\subsection{Conformal flatness}

To guarantee a gauge in which biconformal space is the cotangent bundle
structure of a phase space, we require conformal flatness of the momentum
submanifold. This is accomplished by demanding that the conformal
curvature tensor vanish,\[
C_{b}^{acd}=0\]
where the traceless part of the Riemann curvature, $C_{b}^{acd},$
is given by\begin{eqnarray*}
C_{b}^{acd} & = & R_{b}^{acd}-\frac{1}{n-2}\left(\delta_{b}^{d}R^{ac}-\delta_{b}^{c}R^{ad}+h^{ac}R_{b}^{\quad d}-h^{ad}R_{b}^{\quad c}\right)\\
 &  & +\frac{1}{\left(n-1\right)\left(n-2\right)}R\left(\delta_{b}^{d}h^{ac}-\delta_{b}^{c}h^{ad}\right)\end{eqnarray*}
We make the following claim:
\begin{description}
\item [{Lemma$\;$2:}] Let $h^{ab}$ be symmetric and $h=\eta^{ab}h_{ab}$
an arbitrary function. Then a space with curvature\[
\mathbf{R}_{b}^{a}=-\Delta_{db}^{ac}h^{de}\mathbf{\eta}_{c}\mathbf{\eta}_{e}\]
is conformally flat if and only if $h^{ab}$ has one of the forms\begin{equation}
h^{ab}=\frac{n}{h}\eta^{ab}\label{h conformal to eta}\end{equation}
or\begin{equation}
h^{ab}=\frac{\left(n-2\right)}{hu^{2}}\left(-2u^{a}u^{b}+u^{2}\eta^{ab}\right)\label{Conf flat metric}\end{equation}
for some vector $u^{a}$.
\item [{Proof:}] Expanding $R_{b}^{acd}$ as given by eq.(\ref{Submanifold curvature})
in components and substituting into the expression for vanishing conformal
curvature yields\begin{eqnarray*}
0 & = & -\delta_{b}^{d}h^{ac}+h^{ad}\delta_{b}^{c}+\eta^{ac}\eta_{eb}h^{ed}-\eta^{ad}\eta_{eb}h^{ec}-\frac{\lambda h}{n-2}\left(h^{ac}h_{bg}\eta^{gd}-h^{ad}h_{bg}\eta^{gc}\right)\\
 &  & -\frac{n\left(n-2\right)+\lambda h^{2}}{\left(n-1\right)\left(n-2\right)}\left(\delta_{b}^{c}h^{ad}-\delta_{b}^{d}h^{ac}\right)+\frac{\lambda h}{n-2}\left(\delta_{b}^{c}\eta^{ad}-\delta_{b}^{d}\eta^{ac}\right)\end{eqnarray*}
where we have defined $h\equiv\eta^{ab}h_{ab}$ and $\eta_{ab}h^{ab}=\lambda h$
follows from eq.(\ref{h for antisym}). Though all traces of $C_{b}^{acd}$
using the metric $h^{ab}$ vanish automatically, we get a nontrivial
condition by contracting with $\eta_{ad}.$ Contracting, raising the
lower index with $\eta^{bd}$, using eq.(\ref{h for antisym}), and
collecting terms, yields\[
0=\left(\lambda h^{2}-\left(n-2\right)^{2}\right)\left(h^{ab}-\frac{1}{n}\lambda h\eta^{ab}\right)\]
so either $h^{ab}=\frac{\lambda h}{n}\eta^{ab},$ or $\lambda h^{2}-\left(n-2\right)^{2}=0.$
The first condition fixes $\lambda=\frac{n^{2}}{h^{2}}$, and is immediately
seen to be sufficient to guarantee vanishing conformal curvature.
We seek a sufficient condition when the second condition holds. Expanding
the second condition, we have\[
\left(\eta_{cd}h^{cd}\right)\left(h_{ab}\eta^{ab}\right)=\left(n-2\right)^{2}\]
Substituting this into full expression for vanishing conformal curvature
and raising an index, we find that the equation factors as\begin{equation}
0=k^{ac}k^{bd}-k^{ad}k^{bc}\label{Factored Weyl}\end{equation}
where $k^{ab}=h^{ab}-\frac{n-2}{h}\eta^{ab}$. Now, since $k^{ab}$
is symmetric, it may be diagonalized. It must have at least one nonzero
eigenvalue, which we may take to be $k^{11}$ without loss of generality.
Letting $a=c=1$ in eq.(\ref{Factored Weyl}),\[
k^{bd}=\frac{k^{1d}k^{1b}}{k^{11}}k^{ab}=\rho u^{a}u^{b}\]
where $u^{a}=k^{1a}$. Contracting with $\eta_{bd}$ and using $\lambda h^{2}=\left(n-2\right)^{2}$
shows that $\rho u^{2}=-\frac{2\left(n-2\right)}{h}$. We note that
$u^{2}\neq0$. Re-expressing this result in terms of the the metric
and its inverse yields the forms\begin{eqnarray*}
h^{ab} & = & \frac{n-2}{hu^{2}}\left(-2u^{a}u^{b}+u^{2}\eta^{ab}\right)\end{eqnarray*}
This is the sufficient condition we were seeking, completing the proof
of Lemma 2.
\end{description}
\bigskip{}

This result shows that there are only two possible forms for the induced
metric on configuration space, given the conditions of conjugacy,
orthogonality, and flatness of the momentum submanifold. It is important
to stress that these are the \emph{only} allowed forms of the conformally
flat submanifold metric. The only remaining condition to impose is
involution of the configuration and momentum submanifolds. We end
this Section by showing that the purely conformal case, $h^{ab}=\frac{n}{h}\eta^{ab}$,
does not lead to involute bases.
\begin{description}
\item [{Lemma$\;$3:}] The orthogonal, canonical basis
\end{description}
\begin{eqnarray*}
\mathbf{\chi}^{a} & = & \mathbf{\omega}^{b}+\frac{n}{2h}\eta^{bc}\mathbf{\omega}_{c}\\
\mathbf{\eta}_{a} & = & \frac{1}{2}\mathbf{\omega}_{a}-\frac{h}{n}\eta_{ab}\mathbf{\omega}^{b}\end{eqnarray*}
is \emph{not} involute for any choice of the conformal factor $h$.
\begin{description}
\item [{Proof:}] Beginning with the structure equations for the canonical,
orthogonal basis forms, eqs.(\ref{Chi}) and (\ref{Eta}), we impose
the involution conditions, eqs.(\ref{Chi Involution}) and (\ref{Eta Involution}).
Computing the covariant derivative, $\mathbf{D}h_{ab}=\mathbf{d}h_{ab}-h_{cb}\omega_{a}^{c}-h_{ac}\omega_{b}^{c}+2\omega h_{ab}$,
when metric is given by $h_{ab}=\frac{n}{h}\eta_{ab}$, we find\begin{eqnarray*}
\mathbf{D}h_{ab} & = & -\frac{n}{h^{2}}\left(\mathbf{d}h-2\boldsymbol{\omega}h\right)\eta_{ab}\end{eqnarray*}
where we have used $\eta_{cb}\boldsymbol{\omega}_{a}^{c}+\eta_{ac}\boldsymbol{\omega}_{b}^{c}=0$.
Rewriting the first involution condition in terms of $\mathbf{D}h_{ab}$,
the two conditions may be written as\begin{eqnarray*}
\frac{n}{h^{2}}\left.\left(\mathbf{d}h-2\boldsymbol{\omega}h\right)\mathbf{\eta}_{b}\right|_{\mathbf{\chi}^{d}=0} & = & 0\\
\frac{n}{h^{2}}\left.\left(\mathbf{d}h-2\boldsymbol{\omega}h\right)\mathbf{\chi}^{b}\right|_{\eta_{c}=0} & = & 0\end{eqnarray*}
But if we expand the common factor as $\mathbf{d}h-2\boldsymbol{\omega}h=A_{a}\mathbf{\chi}^{a}+B^{a}\mathbf{\eta}_{a}$
these equations require the vanishing of both $A_{a}$ and $B^{a}$,
and therefore, $\mathbf{d}h-2\boldsymbol{\omega}h=0$. This, in turn,
means the Weyl vector must be exact,\[
\boldsymbol{\omega}=\mathbf{d}\left(\frac{1}{2}\ln h\right)\]
and this is not the case since $\mathbf{d}\boldsymbol{\omega}$ is
the non-degenerate symplectic form. Therefore, the involution conditions
cannot both be satisfied.
\item [{\bigskip{}
}]~
\end{description}
We therefore must restrict our attention to metrics of the form

\[
h_{ab}=\frac{h}{\left(n-2\right)u^{2}}\left(-2u_{a}u_{b}+u^{2}\eta_{ab}\right)\]
This form clearly involves a change of signature between the original
space and the configuration manifold. The remaining Sections are concerned
with this sole remaining possibility.

\section{Solution to the structure equations}

While we have imposed all conditions of Lemma 1 except the involution
condition, we cannot conclude that we have an orthogonal, canonical
basis until we show that all of the structure equations are satisfied.
More importantly, by directly solving the structure equations with
a general orthogonal, canonical basis we establish uniqueness of the
submanifold metric.

The goal in this section, therefore, is to completely solve the structure
equations for the orthogonal canonical basis when the metric is of
the form required by conformal flatness of the momentum submanifold.
The strategy is to first solve for the connection on the momentum
subspace. Since this leads to a different expression for the metric,
we must reconcile the two forms of the metric, and this involves solving
certain coupled differential equations. The second stage of the strategy
is to extend the momentum space form of the connection to the full
biconformal space, then substitute systematically into the full biconformal
structure equations. This procedure yields a complete coordinate expression
for both the connection and the metric, where the metric is of the
required form.

\subsection{The connection of momentum space}

In this next Section, we use the constrained forms of the metric and
connection that we have found to solve the structure equations on
the momentum submanifold. The structure equations are given in equations
eqs.(\ref{Reduced Lorentz}-\ref{Reduced Dilatation}). From eq.(\ref{Reduced Dilatation})
we see immediately that we can gauge the Weyl vector to zero, and
since we know that the space is conformally flat, we can choose the
basis form $\mathbf{\boldsymbol{\eta}}_{a}$ to be conformally exact,
$\mathbf{\boldsymbol{\eta}}_{a}=e^{\sigma}\mathbf{d}y_{a}$. Eq.(\ref{Reduced Eta})
for $\mathbf{\boldsymbol{\eta}}_{a}$ is then solved for the connection,
$\boldsymbol{\beta}_{b}^{a}$, in terms of the gradient of the conformal
function, giving $\mathbf{\beta}_{b}^{a}=-2\Delta_{db}^{ac}\sigma^{,d}\mathbf{d}y_{c}$.
Finally, substituting this expression into the final structure equation
gives a relationship between the conformal function, $\sigma$, and
the metric, $h^{ab}$. Solving this condition, the metric is given
by\[
h^{ac}=e^{-2\sigma}\left(-2\sigma^{,ac}+2\sigma^{,c}\sigma^{,a}-\bar{\sigma}^{2}\eta^{ac}\right)\]
where $\bar{\sigma}^{2}=\eta_{ab}\sigma^{,a}\sigma^{,b}$. The overbar
is to avoid confusion with the square of the function $\sigma$.

Finally, conformal transformation to an exact basis puts the connection
in the form

\begin{eqnarray}
\hat{\boldsymbol{\beta}}_{b}^{a} & = & -2\Delta_{db}^{ac}\sigma^{,d}\mathbf{d}y_{c}\nonumber \\
\hat{\boldsymbol{\eta}}_{a} & = & \mathbf{d}y_{a}\nonumber \\
\hat{\boldsymbol{\tau}} & = & \mathbf{d}\sigma\label{Connection exact gauge}\end{eqnarray}
This form is easily checked by direct substitution. In this exact
gauge, the metric takes the form $\hat{h}^{ac}=-2\sigma^{,ac}+2\sigma^{,c}\sigma^{,a}-\bar{\sigma}^{2}\eta^{ac}$.

\subsection{Reconciling the two forms of the metric}

In Section 4.2, we found that the subspace will be a conformally flat
manifold if and only if\[
h^{ab}=\frac{\left(n-2\right)}{hu^{2}}\left(-2u^{a}u^{b}+u^{2}\eta^{ab}\right)\]
for some vector $u^{a}.$ At the same time, we showed that the metric
must be expressible in the form\[
\hat{h}^{ab}=-2\sigma^{,ab}+2\sigma^{,b}\sigma^{,a}-\bar{\sigma}^{2}\eta^{ab}\]
in the exact basis.

To relate the two forms, we first derive the relationship between
$h$ and $\sigma$. On the momentum submanifold, the integrability
condition reduces to\begin{eqnarray*}
\mathbf{D}h^{ab} & = & \mathbf{d}h^{ab}+h^{cb}\mathbf{\omega}_{c}^{a}+h^{ac}\mathbf{\omega}_{c}^{b}-2h^{ab}\mathbf{\omega}=0\end{eqnarray*}
Using $\eta_{ab}h^{ab}=\frac{1}{h}\left(n-2\right)^{2}$, contraction
with $\eta_{ab}$ gives\begin{eqnarray*}
0 & = & \mathbf{d}h+2h\mathbf{\omega}\end{eqnarray*}
Then with $\mathbf{\omega}=\hat{\boldsymbol{\tau}}=\mathbf{d}\sigma$,
integration immediately gives\begin{eqnarray*}
\frac{1}{h} & = & \pm Ae^{2\sigma}\end{eqnarray*}
for some positive constant $A$.

Replacing $h$ and equating the two resulting forms of $h^{ab},$\[
-2\sigma^{,ab}+2\sigma^{,b}\sigma^{,a}-\bar{\sigma}^{2}\eta^{ab}=\pm\frac{A}{u^{2}}\left(n-2\right)e^{2\sigma}\left(-2u^{a}u^{b}+u^{2}\eta^{ab}\right)\]
We define a new vector, $\hat{u}^{a}=\sqrt{\frac{A\left(n-2\right)}{u^{2}}}e^{\sigma}u^{a}$
with $\hat{u}^{2}=A\left(n-2\right)e^{2\sigma}$. Then\[
-2\sigma^{,ab}+2\sigma^{,b}\sigma^{,a}-\bar{\sigma}^{2}\eta^{ab}=\beta\left(-2\hat{u}^{a}\hat{u}^{b}+\hat{u}^{2}\eta^{ab}\right)\]
where $\beta=\pm1.$ In addition, we have $\hat{u}_{b}\hat{u}^{b,a}=\hat{u}^{2}\sigma^{,a}$,
where $\hat{u}_{b}=\eta_{bc}\hat{u}^{c}.$ To solve this completely,
we also need the integrability condition. Multiplying by a basis form
$\mathbf{d}y_{b}$ we have

\[
\mathbf{d}\sigma^{,a}=\sigma^{,a}\sigma^{,b}\mathbf{d}y_{b}-\frac{1}{2}\bar{\sigma}^{2}\mathbf{d}y^{a}-\beta u^{a}u^{b}\mathbf{d}y_{b}+\beta\frac{1}{2}u^{2}\mathbf{d}y^{a}\]
where $\mathbf{d}y^{a}\equiv\eta^{ab}\mathbf{d}y_{b}$ and we have
dropped the hats on $\hat{u}_{b}$. Taking the exterior derivative
and substituting for all resulting second derivatives, $\sigma^{,ab}$,
we have\begin{eqnarray*}
0 & = & u^{a}u^{b}\sigma^{,c}-u^{a}u^{c}\sigma^{,b}+u^{2}\eta^{ac}\sigma^{,b}-u^{2}\eta^{ab}\sigma^{,c}\\
 &  & +\left(\sigma_{d}u^{d}\right)\eta^{ab}u^{c}-\left(\sigma_{d}u^{d}\right)\eta^{ac}u^{b}+u^{c}u^{a,b}-u^{b}u^{a,c}\\
 &  & +u^{a}u^{c,b}-u^{a}u^{b,c}+u_{d}u^{d,c}\eta^{ab}-u_{d}u^{d,b}\eta^{ac}\end{eqnarray*}
Contraction with $u_{a}$ shows that the curl of $u^{a}$ vanishes
so that $u^{a}=u^{,a}.$ Finally, using this, and simplifying with
$\hat{u}_{b}\hat{u}^{b,a}=\hat{u}^{2}\sigma^{,a}$, we contract with
$u_{c},$ and solve for the second derivative,\[
u^{,ab}=u^{,a}\sigma^{,b}+u^{,b}\sigma^{,a}-\left(u_{c}\sigma^{,c}\right)\eta^{ab}\]
This necessary condition is readily checked to be sufficient as well.

We now have a pair of coupled equations,\begin{eqnarray*}
\sigma^{,ab} & = & \sigma^{,b}\sigma^{,a}+\beta u^{,a}u^{,b}-\frac{1}{2}\left(\bar{\sigma}^{2}+\beta u^{2}\right)\eta^{ab}\\
u^{,ab} & = & u^{,a}\sigma^{,b}+u^{,b}\sigma^{,a}-\left(u_{c}\sigma^{,c}\right)\eta^{ab}\end{eqnarray*}
with the metric given by either of the original forms. Adding $\pm\sqrt{\beta}$
times the second equation to the first decouples these,\begin{eqnarray}
0 & = & \kappa_{\pm}^{,ab}-\kappa_{\pm}^{,a}\kappa_{\pm}^{,b}+\frac{1}{2}\bar{\kappa}_{\pm}^{2}\eta^{ab}\label{Kappa}\end{eqnarray}
where\begin{eqnarray*}
\kappa_{\pm} & = & \sigma\pm\sqrt{\beta}u\end{eqnarray*}
and $\bar{\kappa}_{\pm}^{2}$ is the squared gradient of $\bar{\kappa}_{\pm}$.
Notice that when $\beta=-1,$ these variables are complex, though
$\sigma$ and $u$ remain real. Observe as well that when $\beta=+1,$
we may have either $\kappa_{+}=0$ or $\kappa_{-}=0,$ though not
both at once.

To solve either of eqs.(\ref{Kappa}), first contract with $\kappa_{a}^{\pm}=\eta_{ab}\kappa_{\pm}^{,b}$
and integrate, to find $\kappa_{\pm}^{2}=Ae^{\kappa_{\pm}}$. Substituting
this for $\kappa_{\pm}^{2}$, the result is immediately integrated
twice to give \begin{eqnarray*}
\kappa_{\pm} & = & a_{\pm}-\ln\left(\left(y^{a}+c_{\pm}^{a}\right)\left(y_{a}+c_{\pm a}\right)\right)\end{eqnarray*}
Recalling the zero solutions, there are therefore four cases:

For $\beta=1$ and either $\kappa_{+}=0$ or $\kappa_{-}=0$, we choose
the $y_{a}$ origin at $c_{a}^{\mp}$. Then\begin{eqnarray*}
\mp u & = & \sigma=\frac{a}{2}-\frac{1}{2}\ln\left\vert y^{2}\right\vert \end{eqnarray*}
When neither $\kappa_{+}$ nor $\kappa_{-}$ vanishes we find\begin{eqnarray*}
\sigma & = & \frac{a+b}{2}-\frac{1}{2}\ln\left(\left(y+\sqrt{\beta}a\right)^{2}\right)-\frac{1}{2}\ln\left(\left(y-\sqrt{\beta}a\right)^{2}\right)\\
u & = & \frac{1}{\sqrt{\beta}}\left(\frac{a-b}{2}-\frac{1}{2}\ln\left(\left(y+\sqrt{\beta}a\right)^{2}\right)+\frac{1}{2}\ln\left(\left(y-\sqrt{\beta}a\right)^{2}\right)\right)\end{eqnarray*}
for $\beta=\pm1$ and $\sqrt{\beta}=+i$ when $\beta=-1$.

In each of these cases, we find the metric, $\hat{h}^{ab}=-2\sigma^{,ab}+2\sigma^{,b}\sigma^{,a}-\bar{\sigma}^{2}\eta^{ab}.$
When one of $\kappa_{\pm}$ vanishes,\[
\hat{h}^{ab}=\frac{1}{\left(y^{2}\right)^{2}}\left(-2y^{a}y^{b}+y^{2}\eta^{ab}\right)\]
while in the remaining two cases,\[
\hat{h}^{ab}=-2\left(r^{,a}-s^{,a}\right)\left(r^{,b}-s^{,b}\right)+\left(r-s\right)^{2}\eta^{ab}\]
where\begin{eqnarray*}
r & = & \frac{1}{2}\ln\left(y+\sqrt{\beta}c\right)^{2}\\
s & = & \frac{1}{2}\ln\left(y-\sqrt{\beta}c\right)^{2}\end{eqnarray*}
In every case, the metric is of the general form\begin{equation}
\hat{h}^{ab}=\beta\left(-2u^{a}u^{b}+u^{2}\eta^{ab}\right)\label{Generic metric}\end{equation}
This completes the description of the allowed momentum space solutions.

\subsection{Solving for the full biconformal connection}

In this Section, we extend the solution for the connection on the
momentum submanifold to a form valid on the full biconformal space,
then complete our solution by substituting these forms into the structure
equations, eqs.(\ref{Chi/Eta Lorentz}-\ref{Chi/Eta Weyl}).

On the momentum submanifold, we have the connection in the exact form
given in eqs.(\ref{Connection exact gauge}), the metric given by
eq.(\ref{Generic metric}) where $u^{,a}$ is given by any of the
solutions in the previous Subsection, and $\beta=\pm1.$ Dropping
the circumflex on the exact-basis connection and corresponding metric,
we extend eqs.(\ref{Connection exact gauge}) back to the full manifold,
by adding arbitrary dependence on $\mathbf{\chi}^{a}$ or $\mathbf{d}v^{\beta}$
to each of the connection forms, \begin{eqnarray}
\mathbf{\omega}_{b}^{a} & = & \mathbf{\alpha}_{b}^{a}-2\Delta_{db}^{ac}\sigma^{,d}\mathbf{d}y_{c}\label{Extended Lorentz}\\
\mathbf{\chi}^{a} & = & \chi_{\beta}^{\quad a}\mathbf{d}v^{\beta}\label{Extended Chi}\\
\mathbf{\eta}_{a} & = & \mathbf{d}y_{a}+b_{a\beta}\mathbf{d}v^{\beta}\label{Extended Eta}\\
\mathbf{\omega} & = & W_{\beta}\mathbf{d}v^{\beta}+\sigma^{,b}\mathbf{d}y_{b}\label{Extended Weyl}\end{eqnarray}
where $\mathbf{\alpha}_{b}^{a}=\alpha_{b\beta}^{a}\mathbf{d}v^{\beta}$
and, without loss of generality, we choose the coordinates $v^{\beta}$
canonically conjugate to $y_{\alpha}.$ The coefficients of the new
$\mathbf{d}v^{\beta}$ terms depend arbitrarily on all of the coordinates,
$\left(v^{\beta},y_{\alpha}\right),$ and all of the constants (with
respect to $y_{a}$) in the expressions for $u$ are now allowed to
depend on $v^{a}$, i.e., $\left(a,b,c^{a}\right)\rightarrow\left(a\left(v\right),b\left(v\right),c^{a}\left(v\right)\right)$.

The form of the connection in eqs.(\ref{Extended Lorentz}-\ref{Extended Weyl})
satisfies the conditions of Lemma 1. In order to provide a description
of a biconformal geometry, they must also satisfy the structure equations,
eqs.(\ref{Chi/Eta Lorentz}-\ref{Chi/Eta Weyl}). We begin, in Subsection
5.3.1, with the dilatation equation, eq.(\ref{Chi/Eta Weyl}). At
the start of Subsection 5.3.2, we digress to compute the metric derivatives
required for our discussion of the eta and chi equations and check
the involution conditions. Then, in Subsection 5.3.3 we solve the
remaining structure equation, eqs.(\ref{Chi/Eta Lorentz} - \ref{Eta}).

\subsubsection{Dilatation}

Since $v^{\beta}$ is chosen to be canonically conjugate to $y_{\beta}$
we can write the dilatation equation, eq.(\ref{Chi/Eta Weyl}), in
two ways, $\mathbf{d\omega}=\mathbf{d}v^{\beta}\mathbf{d}y_{\beta}=\mathbf{\chi}^{a}\mathbf{\eta}_{a}$,
leading to two separate conditions. Substituting the form of $\mathbf{\omega}$
from eq.(\ref{Extended Weyl}) into the first gives\begin{eqnarray*}
W_{\left[\alpha,\beta\right]} & = & 0\\
W_{\beta}^{\quad,\alpha}-\sigma_{\quad,\beta}^{,\alpha} & = & -\delta_{\beta}^{\alpha}\end{eqnarray*}
Integrating these immediately gives $W_{\beta}=-y_{\beta}+\sigma_{,\beta}$
where an additional possible function of $v^{\alpha}$ has been absorbed
into the still undetermined $v^{\beta}$-dependent part of $\sigma.$
Substituting into the $\mathbf{d}v^{\beta}\mathbf{d}y_{\beta}=\mathbf{\chi}^{a}\mathbf{\eta}_{a}$,
we see that $b_{\alpha\beta}=\chi_{\beta}^{\quad a}b_{a\beta}$ must
be symmetric and $\chi_{\beta}^{\quad a}=\delta_{\beta}^{a}.$ The
basis form $\mathbf{\chi}^{a}=\delta_{\beta}^{a}\mathbf{d}v^{\beta}=\mathbf{d}v^{a}$
is now exact so there is no longer any need for Greek indices -- the
coordinate basis is also orthonormal. The Weyl vector and $\mathbf{\chi}^{a}$
are now\begin{eqnarray}
\mathbf{\chi}^{a} & = & \mathbf{d}v^{a}\label{Exact Chi}\\
\mathbf{\omega} & = & -y_{a}\mathbf{d}v^{a}+\mathbf{d}\sigma\label{Integrated Weyl}\end{eqnarray}

\subsubsection{Covariant derivative of the metric and the involution conditions}

In order to satisfy the chi and eta structure equations, eqs.(\ref{Chi})
and (\ref{Eta}), we must first evaluate the covariant derivative
of the metric and satisfy the involution condition of Lemma 1.

To simplify the notation, we drop the hat on the metric, $\hat{h}^{ab}\rightarrow h^{ab}$.
While the covariant derivative of $h^{ab}$ vanishes on the involute
submanifolds where (subject to the consistency conditions of Section
6) $h^{ab}$ functions as a metric, the covariant derivative does
not vanish on the biconformal space as a whole. Defining the $\alpha$-
and dilatation-covariant derivative of $u^{a}$ and $h^{ab}$ by\begin{eqnarray*}
u_{\quad;c}^{,a} & \equiv & u_{\quad,c}^{,a}+u^{,d}\alpha_{dc}^{a}-u^{,a}\left(-y_{c}+\sigma_{,c}\right)\\
h_{\quad;c}^{ab} & = & h_{\quad,c}^{ab}+h^{db}\alpha_{dc}^{a}+h^{ad}\alpha_{dc}^{b}-2h^{ab}\left(-y_{c}+\sigma_{,c}\right)\end{eqnarray*}
and expanding\begin{eqnarray*}
\mathbf{D}h^{ab} & = & \mathbf{d}h^{ab}+h^{cb}\mathbf{\omega}_{c}^{a}+h^{ac}\mathbf{\omega}_{c}^{b}-2h^{ab}\mathbf{\omega}\end{eqnarray*}
we find that $\mathbf{D}h^{ab}=h_{\quad;c}^{ab}\mathbf{d}v^{c}$where
$h^{ab}$ is given by eq.(\ref{Generic metric}) and the allowed forms
for $u$ are given by the solutions to eqs.(\ref{Kappa}).

This expression for the covariant derivative of the metric must satisfy
the two involution conditions, eqs.(\ref{Chi Involution}) and (\ref{Eta Involution}),
which now take the form\begin{eqnarray*}
\left.\mathbf{D}h^{ab}\right\vert _{dy}\mathbf{d}y_{b} & = & 0\\
h_{bc}\left.\mathbf{D}h^{ab}\right\vert _{dv}\mathbf{d}v^{c} & = & 0\end{eqnarray*}
The first of these is immediate, since $\left.\mathbf{D}h^{ab}\right\vert _{dy}=0.$
We may rewrite the second as\[
0=h_{bc}h_{\quad;e}^{ab}\mathbf{d}v^{e}\mathbf{d}v^{c}\]
Antisymmetrizing, expanding, and dropping an overall factor leads
to\begin{eqnarray}
0 & = & u_{c}u_{\quad;e}^{,a}-u_{e}u_{\quad;c}^{,a}+u^{,a}u_{e;c}-u^{,a}u_{c;e}\nonumber \\
 &  & +\frac{1}{2}\left(u^{2}\right)_{;e}\delta_{c}^{a}-\frac{1}{2}\left(u^{2}\right)_{;c}\delta_{e}^{a}\label{Involution for chi}\end{eqnarray}

To find the consequences of this condition, we require three identities.
First, contract with $u_{a}$ to find\begin{equation}
0=\left(u^{2}\right)_{;e}u_{c}-\left(u^{2}\right)_{;c}u_{e}+u^{2}u_{e;c}-u^{2}u_{c;e}\label{Ident 1}\end{equation}
A second contraction, with $u^{,e},$ shows that\begin{equation}
u^{,e}u_{c;e}=\frac{1}{u^{2}}u^{,e}\left(u^{2}\right)_{;e}u_{c}-\frac{1}{2}\left(u^{2}\right)_{;c}\label{Ident 2}\end{equation}

A third identity follows by taking the trace on $ae$ in the original
equation, eq.(\ref{Involution for chi}) and using eq.(\ref{Ident 2})
to simplify, giving\begin{eqnarray}
0 & = & \frac{1}{u^{2}}\left(u^{2}u_{\quad;a}^{,a}-u^{,a}\left(u^{2}\right)_{;a}\right)u_{c}-\frac{1}{2}\left(n-2\right)\left(u^{2}\right)_{;c}\label{Step to 3}\end{eqnarray}
 A further simplification occurs when we contract eq.(\ref{Step to 3})
with $u^{c}$ to show that $u^{c}\left(u^{2}\right)_{;c}=\frac{2u^{2}}{n}u_{\quad;a}^{,a}$.
Substituting this result back into eq.(\ref{Step to 3}) results in
our third identity, \begin{equation}
\left(u^{2}\right)_{;c}=\frac{2}{n}u_{\quad;a}^{,a}u_{c}\label{Ident 3}\end{equation}

Together, eq.(\ref{Ident 1}) and eq.(\ref{Ident 3}) show that the
curl of $u_{a}$ vanishes,\[
0=u_{a;c}-u_{c;a}\]
and this with eq.(\ref{Involution for chi}) yields\[
0=u_{c}u_{\quad;e}^{,a}-u_{e}u_{\quad;c}^{,a}+\frac{1}{n}u_{\quad;b}^{,b}u_{e}\delta_{c}^{a}-\frac{1}{n}u_{\quad;b}^{,b}u_{c}\delta_{e}^{a}\]
Finally, contract with $u^{c}$ and use eqs.(\ref{Ident 2}) and (\ref{Ident 3})
to reduce the result to\begin{equation}
u_{\quad;e}^{,a}=\frac{1}{n}u_{\quad;b}^{,b}\delta_{e}^{a}\label{Ness and suff}\end{equation}
Substitution of the result, eq.(\ref{Ness and suff}), into eq.(\ref{Involution for chi})
shows eq.(\ref{Ness and suff}) to be the necessary and sufficient
condition for the involution of chi.

Finally, substituting eq.(\ref{Ness and suff}) into the derivative
of the metric, we have\begin{equation}
\mathbf{D}h^{ab}=-\frac{2\beta}{n}u_{\quad;d}^{,d}\left(\delta_{c}^{a}u^{,b}+u^{,a}\delta_{c}^{b}-u_{c}\eta^{ab}\right)\mathbf{d}v^{c}\label{Derivative of metric}\end{equation}
In this form, it is easy to confirm that both involution conditions
hold.

\subsubsection{Solving the remaining structure equations}

\paragraph{Chi equation}

Substituting the derivative of $h^{ab}$ from eq.(\ref{Derivative of metric}),
and the connection from eqs.(\ref{Extended Lorentz}, \ref{Exact Chi}
and \ref{Integrated Weyl}), the structure equation for chi, eq.(\ref{Chi}),
becomes

\begin{eqnarray}
0 & = & \alpha_{cd}^{a}\mathbf{d}v^{c}\mathbf{d}v^{d}-y_{b}\mathbf{d}v^{b}\mathbf{d}v^{a}+\sigma_{,b}\mathbf{d}v^{b}\mathbf{d}v^{a}\nonumber \\
 &  & -\frac{\beta}{n}u_{\quad;f}^{,f}\left(\delta_{e}^{a}u^{,c}+u^{,a}\delta_{e}^{c}-u_{e}\eta^{ac}\right)\left(b_{cd}+h_{cd}\right)\mathbf{d}v^{e}\mathbf{d}v^{d}\nonumber \\
 &  & -2\Delta_{dc}^{ae}\sigma^{,d}\mathbf{d}v^{c}\mathbf{d}y_{e}+\sigma^{,b}\mathbf{d}y_{b}\mathbf{d}v^{a}-\frac{\beta}{n}u_{\quad;f}^{,f}\left(\delta_{e}^{a}u^{,c}+u^{,a}\delta_{e}^{c}-u_{e}\eta^{ac}\right)\mathbf{d}v^{e}\mathbf{d}y_{c}\label{Chi eq. expand}\end{eqnarray}

The collected cross-terms may be rearranged as\begin{eqnarray*}
0 & = & \left(\sigma^{,a}+\frac{\beta}{n}u_{\quad;b}^{,b}u^{,a}\right)\delta_{c}^{e}-\eta^{ae}\left(\sigma_{c}+\frac{\beta}{n}u_{\quad;b}^{,b}u_{c}\right)+\left(\sigma^{,e}+\frac{\beta}{n}u_{\quad;b}^{,b}u^{,e}\right)\delta_{c}^{a}\end{eqnarray*}
Then, taking the $ec$ trace, shows immediately that\begin{equation}
\sigma^{,a}=-\frac{\beta}{n}u_{\quad;d}^{,d}u^{,a}\label{u and sigma}\end{equation}
This means that $\sigma^{,a}$ and $u^{,a}$ must be parallel, since
neither can vanish. However, it is straightforward to show that this
can never happen unless either $\kappa_{+}=0$ or $\kappa_{-}=0$,
so we discard the other solutions to eqs.(\ref{Kappa}) and set $\pm u=\sigma=\frac{a\left(v\right)}{2}-\frac{1}{2}\ln\left\vert y^{2}\right\vert $
and $\beta=1.$ This gives the final form of the metric,\begin{equation}
h_{ab}=-\left(2y_{a}y_{b}-y^{2}\eta_{ab}\right)\label{Final metric}\end{equation}

These results have some useful consequences. First, substituting $\sigma^{,a}$
for $\pm u^{,a}$ in eq.(\ref{u and sigma}), yields $\sigma_{\quad;d}^{,d}=-n.$
We may combine this with the result for the covariant derivative of
$u^{a},$ eq.(\ref{Ness and suff}), to show that $\sigma_{\quad;c}^{,a}=-\delta_{c}^{a}$.

Now, expand the covariant derivative, $\sigma_{\quad;c}^{,a}=\sigma_{\quad,c}^{,a}+\sigma^{,b}\alpha_{bc}^{a}-\sigma^{,a}\left(-y_{c}+\sigma_{,c}\right)$.
Substituting $\sigma^{,a}=-\frac{y^{a}}{y^{2}}$ and $\sigma_{,a}=\frac{1}{2}a_{,a},$
and contracting with $y_{a}$ now shows that $\sigma$ is independent
of $v^{a},$\[
\sigma_{,a}=0\]
while we still have the condition\begin{equation}
0=-y^{b}\alpha_{bc}^{a}-y^{a}y_{c}+y^{2}\delta_{c}^{a}\label{Last condition}\end{equation}
We will return to this condition after considering the configuration
space terms ($\mathbf{d}v^{c}\mathbf{d}v^{d}$) of eq.(\ref{Chi eq. expand}).

For the configuration space terms, we may now set $\beta=1,$ $u^{,a}=\pm\sigma^{,a},$
$\sigma_{,c}=0$ and $\sigma_{\quad;d}^{,d}=-n.$ Then, antisymmetrizing
and dropping the basis forms,\begin{eqnarray*}
0 & = & \alpha_{ed}^{a}-\alpha_{de}^{a}-y_{e}\delta_{d}^{a}+y_{d}\delta_{e}^{a}+\left(\sigma^{,c}\delta_{e}^{a}+\sigma^{,a}\delta_{e}^{c}-\sigma_{e}\eta^{ac}\right)b_{cd}\\
 &  & -\left(\sigma^{,c}\delta_{d}^{a}+\sigma^{,a}\delta_{d}^{c}-\sigma_{d}\eta^{ac}\right)b_{ce}\end{eqnarray*}
Lowering $a$ and permuting the three free indices, we add the first
two and subtract the third. Replacing $\sigma^{,a}$ with its explicit
form, this gives an explicit expression for the $\mathbf{d}v^{a}$
part of the spin connection,\begin{eqnarray}
\alpha_{bc}^{a} & = & -y^{a}\eta_{bc}+y_{b}\delta_{c}^{a}-\frac{1}{y^{2}}\left(y_{b}b_{\quad c}^{a}-y^{a}b_{bc}+y^{e}b_{eb}\delta_{c}^{a}-y^{e}b_{\quad e}^{a}\eta_{bc}\right)\label{Alpha}\end{eqnarray}

Returning to the cross-term, we impose the condition, eq.(\ref{Last condition})
on the explicit form for $\alpha_{bc}^{a}$ above. Lowering the $a$
index, and defining $b_{a}=y^{b}b_{ab}$ gives $b_{ac}$ in terms
of $b_{a},$ \begin{equation}
b_{ac}=\frac{1}{y^{2}}\left(b_{a}y_{c}+y_{a}b_{c}-\eta_{ac}\left(y^{b}b_{b}\right)\right)\label{Form of bab}\end{equation}
This completes the consequences of the chi structure equation. We
now turn to the eta equation.

\paragraph{Eta equation}

Substituting the previous results (eqs.(\ref{Chi/Eta Lorentz}), (\ref{Eta}),
(\ref{Exact Chi}), (\ref{Integrated Weyl}), (\ref{Alpha}), (\ref{Form of bab}),
(\ref{Derivative of metric})) into the structure equation for $\mathbf{\eta}_{a}$,
eq.(\ref{Eta}), gives three independent equations,

\begin{eqnarray*}
0 & = & \frac{1}{y^{2}}\left(y^{b}\delta_{a}^{c}-\eta^{bc}y_{a}+y^{c}\delta_{a}^{b}\right)\mathbf{d}y_{c}\mathbf{d}y_{b}\\
0 & = & \left(-b_{bc}^{\quad,a}+\frac{1}{y^{2}}\left(b_{b}\delta_{c}^{a}-b^{a}\eta_{bc}+\delta_{b}^{a}b_{c}\right)\right)\mathbf{d}y_{a}\mathbf{d}v^{c}\\
0 & = & \left(-b_{ae,c}+\alpha_{ac}^{b}b_{be}+y_{c}b_{ae}+\left(\eta_{ac}y^{d}-y_{a}\delta_{c}^{d}-y_{c}\delta_{a}^{d}\right)b_{de}\right)\mathbf{d}v^{c}\mathbf{d}v^{e}\end{eqnarray*}
The first equation, in $\mathbf{d}y_{c}\mathbf{d}y_{b}$, is seen
to vanish identically. The $\mathbf{d}y_{a}\mathbf{d}v^{c}$ cross
term, carrying out the final substitution for $b_{ab}$ and collecting
terms, may be cast into the form\[
0=\left(y^{2}y_{b}\delta_{a}^{d}+y^{2}y_{a}\delta_{b}^{d}-y^{2}\eta_{ab}y^{d}\right)\left(\frac{b_{d}}{y^{2}}\right)^{,c}\]
Contraction with $y^{b}$ now shows that the final factor must vanish,
$\left(\frac{b_{a}}{y^{2}}\right)^{,c}=0$, so we may write $b_{a}$
as $y^{2}$ times a vector dependent only on $v^{a},$ $b_{a}=y^{2}d_{a}\left(v\right)$.
Then $b_{ab}$ takes the form\[
b_{ab}=y_{b}d_{a}+y_{a}d_{b}-\eta_{ab}y^{e}d_{e}\]
A quick check shows that this result is necessary and sufficient to
satisfy the cross-term equation.

Finally, expanding the third equation in terms of $d_{a}\left(v\right)$
yields an expression linear in $y_{a}.$ Differentiating with respect
to $y_{d},$ then taking the $bd$ trace leaves us with\begin{eqnarray*}
0 & = & \left(n-1\right)d_{a,c}-d_{c,a}+\eta_{ac}d_{\quad,b}^{b}-\left(n-2\right)\eta_{ac}\left(d^{e}d_{e}\right)+\left(n-2\right)d_{a}d_{c}\end{eqnarray*}
Taking the antisymmetric part shows that $\frac{n}{2}\left(d_{a,c}-d_{c,a}\right)=0,$
and therefore $d_{a}$ is a gradient, $d_{,a}$, while the contraction
with $\eta^{ac}$ shows that $d_{\quad,b}^{b}=\frac{1}{2}\left(n-2\right)d^{c}d_{c}.$
Substituting gives a differential equation for $d,$\begin{equation}
d_{,ac}=-d_{,a}d_{,c}+\frac{1}{2}\eta_{ac}\left(\eta^{bd}d_{,b}d_{,d}\right)\label{Diff eq for d}\end{equation}
This relationship is easily shown to be necessary and sufficient for
the solution of the full equation.

Eq.(\ref{Diff eq for d}) is straightforward to integrate. First,
contract with $d^{c}.$ Then if $d^{2}=\left(\eta^{bd}d_{,b}d_{,d}\right)$
is nonzero we may integrate to show that $d^{2}=Ae^{-d}.$ Replacing
$d^{2}$ in the original equation allows two immediate integrations,
which produce the result $d=\ln\left(v^{a}+\frac{2A^{a}}{A}\right)^{2}+\ln\left(\frac{A}{4}\right)$
and since changing the origin of the $v^{a}$ coordinate does not
change the conjugacy of $v^{a}$ and $y_{a},$ we may shift the origin
so that\[
d=\ln\left\vert v^{2}\right\vert +c\]
where $c$ is an arbitrary constant. For the $d^{2}=0$ case, integration
leads to the alternative solution\[
d=\ln\left(a+b_{a}v^{a}\right)\]
where $b_{a}b^{a}=0.$ The special case, $a=1,$ $b_{a}=0$ shows
that $d=0$ is a solution.

The $y_{a}=$ constant subspaces coincide with the $\mathbf{\eta}_{a}=0$
submanifolds if and only if $d=$ constant. In this case, both $\mathbf{\chi}^{a}$
and $\mathbf{\eta}_{a}$ are exact. For other solutions for $d,$
the submanifolds are given by solutions of\[
\mathbf{d}y_{a}+\left(y_{b}d_{a}+y_{a}d_{b}-\eta_{ab}\left(y^{e}d_{e}\right)\right)\mathbf{d}v^{b}=0\]
for functions $y_{a}\left(v^{b}\right).$

With these results, rewritten in the original $\left(\mathbf{\chi}^{a},\mathbf{\eta}_{a}\right)$
basis, we have completely determined the connection:\begin{eqnarray}
\mathbf{\omega}_{b}^{a} & = & 2\Delta_{db}^{ac}\left(y_{c}-d_{c}\right)\mathbf{\chi}^{d}+\frac{2}{y^{2}}\Delta_{db}^{ac}y^{d}\mathbf{\eta}_{c}\label{Final spin connection}\\
\mathbf{\chi}^{a} & = & \mathbf{d}v^{a}\label{Final form basis}\\
\mathbf{\eta}_{a} & = & \mathbf{d}y_{a}+\left(y_{b}d_{a}+y_{a}d_{b}-\eta_{ab}\left(y^{e}d_{e}\right)\right)\mathbf{d}v^{b}\label{Final co-form basis}\\
\mathbf{\omega} & = & \left(-y_{a}+d_{a}\right)\mathbf{\chi}^{a}-\frac{1}{y^{2}}y^{a}\mathbf{\eta}_{a}\label{Final connection}\end{eqnarray}
where $d_{a}=d_{,a}$ and\begin{eqnarray}
d & = & \left\{ \begin{array}{cc}
\ln\left\vert v^{2}\right\vert +c_{0} & d^{2}\neq0\\
\ln\left\vert a_{0}+c_{a}v^{a}\right\vert  & d^{2}=0\end{array}\right.\label{Solution for d}\end{eqnarray}

These results, eqs.(\ref{Solution for d}) and eqs.(\ref{Final spin connection}-\ref{Final connection}),
together with the final form of the metric, eq.(\ref{Final metric}),
establish the form of the connection claimed in Section 2.3.

\paragraph{Spin connection}

Finally, we must check the structure equation for the spin connection,
eq.(\ref{Chi/Eta Lorentz}). Substituting the connection given in
eqs.(\ref{Final spin connection}-\ref{Final connection}) with $d$
given by eq.(\ref{Solution for d}) leads to three extremely long
equations, but each equation is identically satisfied. We have therefore
satisfied the structure equations, and all conditions of Lemma 1.

We complete our proof of the Signature Theorem by checking the signature
of $h_{ab}.$

\section{The existence of time}

Even though our candidate submanifold metric, $h_{ab}$ or $-h^{ab}$,
is now a uniquely specified, invertible, symmetric quadratic form,
its signature is not consistent for every choice of the original metric
$\eta_{ab}$. The inconsistencies take different forms in the two
cases. For the configuration submanifolds, the signature of $-h^{ab}$
is consistent on any one configuration submanifold, but may vary as
we look at different submanifolds. This means that the signature of
spacetime would change for particles of different momenta, which we
disallow. For the momentum submanifolds, the situation is even worse:
$h_{ab}$ may have inconsistent signature on a given momentum submanifold.
The proof of the Signature Theorem requires consistent signature for
the induced metric on at least one of the classes of submanifold,
configuration or momentum. Either class yields the same conclusion.

In this Section, we find the conditions under which these metrics
are assured to have consistent signature. The full biconformal line
element is given by expanding in the orthonormal basis, $ds^{2}=h_{ab}\chi^{a}\chi^{b}-h^{ab}\eta_{a}\eta_{b}$,
then expanding the basis in terms of coordinate differentials,\begin{eqnarray*}
ds^{2} & = & \left(h_{cd}-h^{ab}b_{ac}b_{bd}\right)dv^{c}dv^{d}-2h^{cb}b_{bd}dy_{c}dv^{d}-h^{ab}dy_{a}dy_{b}\end{eqnarray*}
where $b_{ab}=y_{b}d_{a}+y_{a}d_{b}-\eta_{ab}\left(y^{e}d_{e}\right)$
and $d_{a}=d_{,a}$ is given in eq.(\ref{Solution for d}). We consider
first the configuration spaces $\left(\mathbf{\eta}_{b}=0\right)$,
then the momentum spaces $\left(\mathbf{\chi}^{a}=0\right).$

\subsection{Configuration submanifold}

Setting $\mathbf{\eta}_{a}=0$ requires a relationship between the
$y_{a}$ and v$^{a}$ coordinates, with $y_{a}\left(v\right)$ found
by solving $dy_{a}=-b_{ab}dv^{b}$. While the solution is straightforward,
we need only this relationship. From $dy_{a}=-b_{ab}dv^{b}$ we must
have $\frac{\partial y_{a}}{\partial v^{b}}=-b_{ab}$, and since $b_{ab}$
is symmetric, $y_{a}$ must be a gradient, $y_{a}=t_{,a}$, of some
function $t\left(v^{a}\right)$. Taking $t$ as a coordinate, we have
$y_{a}dv^{a}=dt.$ Substituting this expression and $dy_{a}=-b_{ab}dv^{b}$
into the full line element reduces it to simply\[
ds^{2}=y^{2}\eta_{ab}dv^{a}dv^{b}-2dt^{2}.\]
Project $dv^{a}$ into parts parallel and orthogonal to $dt,$ $\mathbf{d}v^{a}=\mathbf{e}_{\parallel}^{a}+\mathbf{e}_{\perp}^{a}$,
and the line element becomes\begin{eqnarray*}
ds^{2} & = & y^{2}\eta_{ab}\mathbf{e}_{\perp}^{a}\mathbf{e}_{\perp}^{b}-dt^{2}\end{eqnarray*}
The signature of this line element depends on the sign of $y^{2},$
which changes as $v^{a}$ varies over the submanifold and as $\left(v^{a},y_{b}\right)$
vary over the full biconformal space. Note that these arguments hold
only if both signs of $y^{2}$ occur, so if $\eta_{ab}$ is Euclidean,
with signature $s_{0}=\pm n$, then $y^{2}\eta_{ab}\mathbf{e}_{\perp}^{a}\mathbf{e}_{\perp}^{b}$
is positive definite. In these Euclidean cases, the submanifold signature
is $s=n-2$, and $h_{ab}$ is consistently Lorentzian.

Now suppose $\eta_{ab}$ has signature different from $n.$ Then depending
on the form of $y_{a}\left(v\right)$, it may be possible to have
$y^{2}$ either positive (spacelike) or negative (timelike) on a given
submanifold, and as we look at distinct submanifolds at different
points of the full biconformal space, $y^{2}$ must change sign. Let
the signature of the full metric be $s_{0}=p-q,$ and suppose $y^{2}\eta_{ab}\mathbf{e}_{\perp}^{a}\mathbf{e}_{\perp}^{b}$
has signature $s^{\prime}.$ Then when $y_{a}$ is spacelike, $y^{2}>0$,
the metric $h_{ab}$ has signature $s=s^{\prime}-1.$ Also, since
the basis $\mathbf{e}_{\perp}^{a}$ is orthogonal to $y_{a}$ and
$y_{a}$ is spacelike, $s'$ must equal $s_{0}-1$. Putting these
together we see that $s=s_{0}-2$.

When $y^{2}$ is timelike, the argument is similar. We still have
$s=s'-1$, but this time the signature of $\eta_{ab}\mathbf{e}_{\perp}^{a}\mathbf{e}_{\perp}^{b}$
is $s_{0}+1$. Therefore, with $y^{2}$ negative, the signature of
$y^{2}\eta_{ab}\mathbf{e}_{\perp}^{a}\mathbf{e}_{\perp}^{b}$ is $s'=-\left(s_{0}+1\right)$.
Combining gives $s=-s_{0}-2$. 

Demanding consistent signature, $s=s_{0}-2=-s_{0}-2$, we can only
have vanishing initial signature, $s_{0}=0$.

We must still check whether $y^{2}$ actually does change sign on
the configuration submanifold, where $y_{a}$ is a function of $v^{a}.$
Integrating to find $y_{a}\left(v^{b}\right)$ gives the generic solution,
$y_{a}=\frac{1}{\left(v^{2}\right)^{2}}\left(v^{2}\delta_{a}^{c}-2v_{a}v^{c}\right)a_{c}$
in which case $y^{2}=\frac{1}{\left(v^{2}\right)^{2}}\eta^{ab}a_{a}a_{b}$
so the submanifold at any given point $\left(v^{a},y_{b}\right)$
has consistent signature. However, as we look at distinct configuration
submanifolds, we require different vectors $a_{c}$ in order to get
all possible vectors $y_{a.}$ Therefore, for the generic solution
for $y_{a}\left(v^{b}\right),$ distinct configuration submanifolds
will have different signature unless $s_{0}=0$ or $s_{0}=\pm n.$
There is a special case in which $y_{a}$ is constant, and the same
consideration applies.

There is one final special solutions for $y_{a}\left(v\right)$ of
the form $y_{a}=\frac{c}{A}v^{a}$. In this case $y^{2}$ has the
same sign as $v^{2}$, so as $v^{a}$ varies over the (non-Euclidean)
submanifold, $v^{2}$ will change sign. Then the signature is not
consistent even on a fixed submanifold. 

We conclude that while for any signature, $s_{0},$ of $\eta_{ab},$
there exist solutions for the configuration submanifolds such that
each submanifold has consistent signature, configuration submanifolds
at different points of the biconformal space (hence different momenta)
have different signatures unless $s_{0}=\pm n$ or $s_{0}=0.$

\subsection{Momentum submanifold}

When $\mathbf{\chi}^{a}=0$, we have $v^{a}=$ constant. The candidate
metric is $-h^{ab}$ and the line element is\[
ds^{2}=-\frac{1}{\left(y^{2}\right)^{2}}\left(-2\left(y^{a}dy_{a}\right)\left(y^{b}dy_{b}\right)+y^{2}\eta^{ab}dy_{a}dy_{b}\right)\]
When $v^{a}=$ constant, the $y_{a}$ coordinates are no longer all
independent, since the length of $y_{a}$ in the induced candidate
metric is $\left\vert y\right\vert ^{2}=-h^{ab}y_{a}y_{b}=1$. However,
$y^{2}=\eta^{ab}y_{a}y_{b},$ remains independent, so we choose $y^{2}$
together with the $y_{a}$ to span the space. We again consider positive
and negative values for $y^{2}$. 

When $y^{2}>0$, define $r^{2}=y^{2}>0$ so that the line element
becomes\begin{eqnarray*}
ds^{2} & = & \frac{1}{r^{2}}\left(2dr^{2}-\eta^{ab}dy_{a}dy_{b}\right)\end{eqnarray*}
Now project $dy_{a}$ into parts along $dr$, given by $e_{a}^{\Vert}=\frac{y_{a}}{r}dr$,
and parts perpendicular, $e_{a}^{\bot}=dy_{a}-\frac{y_{a}}{r}dr$.
Then defining $t=\ln r$ the line element becomes\[
ds^{2}=dt^{2}-e^{2t}\eta^{ab}e_{b}^{\bot}e_{a}^{\bot}\]
Since $y^{2}>1,$ the quadratic form $e^{2t}\eta^{ab}e_{b}^{\bot}e_{a}^{\bot}$
has signature $s^{\prime}=s_{0}-1,$ so that the metric of the momentum
space is of signature $s=1-\left(s_{0}-1\right)=-s_{0}+2$.

For $y^{2}<0$ we define $r^{2}=-y^{2}>0$. This time the line element
becomes\[
ds^{2}=\frac{1}{r^{2}}\left(2dr^{2}+\eta^{ab}dy_{a}dy_{b}\right)\]
Projecting $dy_{a}$ into parts along $dr$ and perpendicular to $dr$
as before, $\eta^{ab}dy_{a}dy_{b}=\eta^{ab}\left(\frac{y_{a}y_{b}}{r^{2}}dr^{2}+e_{a}^{\bot}e_{b}^{\bot}\right)$
so we have, with $t=\ln r$,

\[
ds^{2}=dt^{2}+\frac{1}{r^{2}}\eta^{ab}e_{a}^{\bot}e_{b}^{\bot}\]
Again consider the signature of the candidate metric. Since the perpendicular
part has lost a timelike piece, $\frac{1}{r^{2}}\eta^{ab}e_{a}^{\bot}e_{b}^{\bot}$
has signature $s=s_{0}+1.$ Including the positive $dt^{2}$ makes
the total signature $s=s_{0}+2$.

Comparing results, we once again find that $s_{0}=0$ is the only
non-Euclidean solution. The Euclidean cases work as before, with $s=-s_{0}+2$
for $s_{0}=+n$ and $s=s_{0}+2$ when $s_{0}=-n$.

Finally, since $y_{a}$ spans the momentum space with $v^{a}$ constant,
the sign of $y^{2}$ will necessarily vary unless $\eta^{ab}$ is
Euclidean, $s_{0}=\pm n.$ Therefore, when $\eta^{ab}$ is non-Euclidean,
the only way to get a consistent signature on \textit{any} fixed submanifold
is to require $s_{0}=0$. For the Euclidean cases, when $s_{0}=\pm n,$
$y^{2}$ is always positive or negative definite and the momentum
space signature is Lorentzian, $s=-n+2.$

We conclude that unless the signature, $s_{0},$ of $\eta^{ab}$ is
$s_{0}=\pm n$ or $s_{0}=0,$ there do not exist any momentum submanifolds
of consistent signature.

\section{Discussion}

The biconformal gauging of the conformal group, in which we form the
group quotient of the conformal group, $SO\left(p+1,q+1\right)$ of
$\mathcal{S}_{p,q}^{n}$, by its Weyl subgroup, $W\left(p,q\right)$,
leads to a $2n$-dimensional manifold possessing natural metric and
symplectic structures \emph{arising necessarily from properties of
the conformal group.} The curved manifolds based on this space admit
an action functional linear in the curvature and the resulting field
equations describe $n$-dim general relativity on the co-tangent bundle.
These curved spaces generically retain the metric and symplectic structures.
As a result, the biconformal gauging gives a systematic construction
of phase spaces associated with pseudo-Euclidean spaces, $\mathcal{S}_{p,q}^{n},$
of arbitrary dimension $n=p+q$ and signature $s_{0}=p-q.$

Defining a metric phase space to be a phase space with metric, having
orthogonal configuration and momentum submanifolds, we have proved
the following theorem:

\textbf{Signature Theorem\qquad{}}\textit{A flat} $2n$\textit{-dim
(}$n>2$\textit{) biconformal space,} $B\left(\mathcal{S}_{p,q}^{n}\right)$,\textit{
is a metric phase space if and only if the signature,} $s_{0}=p-q$,\textit{
of} $\mathcal{S}_{p,q}^{n}$\textit{ is} $\pm n$\textit{ or} $0.$\textit{
The resulting configuration submanifold is Lorentzian, $s=n-2$, when
$s_{0}=\pm n$}, \textit{and of signature $s=2$ when $s_{0}=0$}.

In particular, for the description of any $4$-dim geometry or any
odd-dimensional geometry we have the immediate special cases,

\textbf{Corollary 1 (4-dim)\qquad{}}\textit{A flat} $8$\textit{-dim
biconformal space,} $B\left(\mathcal{S}_{p,q}^{4}\right)$,\textit{
is a metric phase space if and only if the signature,} $s,$\textit{
of} $\mathcal{S}_{p,q}^{4}$\textit{ is} $\pm4$\textit{ or} $0.$\textit{
The resulting configuration space is necessarily Lorentzian.}

\textbf{Corollary 2 \qquad{}}\textit{An odd-dimensional biconformal
space,} $B\left(\mathcal{S}_{p,q}^{4}\right)$,\textit{ is a metric
phase space if and only if }$\mathcal{S}_{p,q}^{4}$\textit{ is Euclidean,
$s=\pm n$}.\textit{ The resulting configuration space is necessarily
Lorentzian.}

\medskip{}

A physical interpretation of the theorem is that we may think of the
world as a generalization of Euclidean space, gauged to produce a
gravity theory. The gauging which reproduces general relativity describes
a $2n$-dimensional phase space, so that the full solution describes
the co-tangent bundle of an $n$-dimensional configuration space.
Time arises only when we carefully identify that configuration space
by systematically dividing the phase space into orthogonal configuration
and momentum pieces. The restriction of the Killing metric to the
configuration space then necessarily has Lorentzian signature, hence,
time.

There are a number of potential applications of the theorem, the most
interesting having to do with quantum gravity. Unlike the twistor
space approach to conformal general relativity, which has infinitely
many Euclidean and Lorentzian submanifolds, the natural symplectic
and metric structures of biconformal space allow the Signature Theorem
to establish a nearly unique connection between Euclidean and Lorentzian
spaces. We now see that \emph{time} may be viewed as a descriptive
property of an initial Euclidean space. If we can formulate a consistent
theory of Euclidean quantum gravity on phase space, then the Theorem
specifies uniquely how to view that theory from spacetime. 

It is possible that the relationship between the underlying Euclidean
space, and the emergent spacetime is related to the well-known Wick
rotation. This rotation in the complex time plane replaces the usual
Minkowski time by a periodic imaginary coordinate, giving a $4$-dim
Euclidean space on which quantum theory becomes stochastic. The Feynman
path integral becomes an ordinary Wiener path integral and the new
coordinate acts as a temperature. This approach has allowed a proof
of convergence of path integrals with a wide range of potentials.
In the approach described here, both Euclidean and Minkowski spaces
are present from the start, so it might be possible to realize the
Wick rotation as simply the quantization procedure pulled back to
the underlying Euclidean space. Alternatively, it may be found to
relate distinct submanifolds of the biconformal space.

One consequence of the emergence of the Lorentzian metric $h_{ab}=-2y_{a}y_{b}+y^{2}\eta_{ab}$
is a possible explanation of the absence of tachyons. Since the coordinate
$y_{a}$ is the conjugate momentum to configuration coordinates $v^{a}$,
and it is $y_{a}$ which determines the timelike direction through
$h_{ab}$, it is likely that spacelike momenta are impossible.

These and other questions will be addressed in subsequent work.

\bigskip{}


\begin{thebibliography}{19}
\bibitem{Utiyama} Utiyama, R., Phys.Rev. \textbf{101} (1956) 1597.

\bibitem{Kibble1} Kibble, T.W.B, J.Math.Phys. \textbf{2} (1961) 212-221.

\bibitem{Neeman} Ne'eman, Y. and Regge, T., Gravity And Supergravity
As Gauge Theories On A Group Manifold, Phys.Lett.B \textbf{74} (1978)
54.

\bibitem{ReggeN} Ne'eman, Y. and Regge, T., Gauge Theory Of Gravity
And Supergravity On A Group Manifold, Riv. Nuovo Cim. \textbf{1N5}
(1978), 1.

\bibitem{Kobayashi} Kobayashi, S., Nomizu, K., \textit{Foundations
of Differential Geometry} (Wiley, New York, 1963).

\bibitem{IvanovI} Ivanov, E.A. and Niederle, J., Gauge formulation
of gravitation theories. I. The Poincaré, de Sitter, and conformal
cases, Phys.Rev.D \textbf{25} (1982) 976.

\bibitem{Ivanov} Ivanov, E.A. and Niederle, J., Gauge formulation
of gravitation theories. II. The special conformal case, Phys.Rev.D
\textbf{25} (1982) 988.

\bibitem{Folger}T. Folger, \emph{Newsflash: Time may not Exist},
Discover Magazine, (June, 2007) http://discovermagazine.com/2007/jun

\bibitem{Flanagan}E. Flanagan, \emph{Fourth order Weyl Gravity},
Phys. Rev. D74 (2006) 023002.

\bibitem{JWAux} J. T. Wheeler, Phys Rev D\textbf{44}, 1769 (1991). 

\bibitem{JW} Wheeler, J.T., New conformal gauging and the electromagnetic
theory of Weyl, J.Math.Phys. \textbf{39} (1998) 299-328. {[}arXiv:hep-th/9706214{]}

\bibitem{WW} Wehner, A. and Wheeler, J. T., Nucl. Phys. B \textbf{557}
(1999) 380-406, arXiv:hep-th/9812099.

\bibitem{Berezin} F. Berezin, Sov. Math. Izv. 38 (1974) 1116; Sov.
Math. Izv. 39 (1975) 363.

\bibitem{Woodhouse} N.M.J. Woodhouse, \textit{Geometrical Quantization},
(Clarendon, New York, 1992).

\bibitem{Curtwright} C. Zachos, T. Curtright, Phase space quantization
of field theory, Prog. Theor. Phys. Suppl. 135 (1999) 244-258.

\bibitem{Klauder} J. Klauder, Phase space geometry in classical and
quantum mechanics, arXiv:quant-ph/0112010v1.

\bibitem{AbrahamAndMarsden} Abraham, R.H. and Marsden, J.E., \textit{Foundations
of Mechanics} (Addison-Wesley, 1978).

\bibitem{YQMisC} Wheeler, J. T., \textit{Why quantum mechanics is
complex,} Bull. Astr. Soc. Ind. \textbf{25} (1997) 591-599, Published
by invitation; received Honorable Mention for the 1996 General Relativity
and Gravitation Awards for Essays on Gravitation. {[}arxiv:hep-th/9708088{]}. 

\bibitem{QM} Anderson, L.B., Wheeler, J.T., Quantum mechanics as
a measurement theory on biconformal space, Int.J.Geom.Meth.Mod.Phys.
\textbf{3} (2006) 315, {[}arXiv:hep-th/0406159{]}.
\end{thebibliography}
\end{document}